\documentclass[a4paper,aps,showkeys,showpacs,amsmath,amssymb,groupedaddress,12pt,preprint]{revtex4-1}

\usepackage{amsfonts}
\usepackage{graphicx}
\usepackage{natbib}
\usepackage{dcolumn}
\usepackage{bm}
\usepackage[letterpaper,left=0.8in,right=0.8in,top=0.7in,bottom=0.8in]{geometry}
\usepackage{rotating}
\usepackage{longtable}
\usepackage{subfigure}
\usepackage{lscape}
\usepackage{setspace}

\newcommand{\beq}{\begin{equation}}
\newcommand{\eeq}{\end{equation}}
\newcommand{\bea}{\begin{eqnarray}}
\newcommand{\eea}{\end{eqnarray}}
\newcommand{\ba}{\begin{array}}
\newcommand{\ea}{\end{array}}
\newcommand{\bi}{\begin{itemize}}
\newcommand{\ei}{\end{itemize}}
\newcommand{\ben}{\begin{enumerate}}
\newcommand{\een}{\end{enumerate}}

\renewcommand{\r}{\right}
\renewcommand{\l}{\left}
\long\def\symbolfootnote[#1]#2{\begingroup\def\thefootnote{\fnsymbol{footnote}}\footnote[#1]{#2}\endgroup}

\begin{document}

\preprint{}

\title{Identifying the Community Structure of the International-Trade Multi Network}

\author{Matteo Barigozzi} 
\affiliation{Department of Statistics,\\ 
London School of Economics and Political Science, UK.\\ E-mail: M.Barigozzi@lse.ac.uk}
\author{Giorgio Fagiolo} 
\affiliation{Laboratory of Economics and Management,\\
Sant'Anna School of Advanced Studies, Pisa, Italy.\\ 
E-mail: giorgio.fagiolo@sssup.it}
\author{Giuseppe Mangioni}
\affiliation{Dipartimento di Ingegneria Informatica e delle Telecomunicazioni,\\
University of Catania, Italy.\\
E-mail: gmangioni@diit.unict.it\\ }

\begin{abstract}
\noindent We study the community structure of the multi-network of commodity-specific trade relations among world countries over the 1992-2003 period. We compare structures across commodities and time by means of the normalized mutual information index (NMI). We also compare them with exogenous community structures induced by geographical distances and regional trade agreements. We find that commodity-specific community structures are very heterogeneous and much more fragmented than that characterizing the aggregate ITN. This shows that the aggregate properties of the ITN may result (and be very different) from the aggregation of very diverse commodity-specific layers of the multi network. We also show that commodity-specific community structures, especially those related to the chemical sector, are becoming more and more similar to the aggregate one. Finally, our findings suggest that geographical distance is much more correlated with the observed community structure than regional-trade agreements. This result strengthens previous findings from the empirical literature on trade.
\end{abstract}



\maketitle

\section{Introduction} \label{Sec:Intro}

In the last years there was a surge of interest in the study of
international-trade issues from a complex-network perspective
\citep{LiC03,SeBo03,Garla2004,Garla2005,serrc07,Bhatta2007a,Bhatta2007b,
Garla2007,Fagiolo2008physa,Fagiolo2008acs,Fagiolo2009pre}. 
Many contributions have  explored the evolution over time of the
topological properties of the aggregate International Trade Network (ITN), aka
the World Trade Web (WTW), defined as the graph of total import/export
relationships between world countries in a given year. More recently, a number of papers have instead begun to investigate the multi-network of trade \citep{ReichardtWhite2007,Barigozzi_etal2010pre}, where a commodity-specific approach
is followed to unfold the aggregate ITN in many layers, each one representing import and
export relationships between countries for a given commodity class (cf. also Refs. \cite{Hidalgo_etal_2007_Science,Hidalgo_Hausmann_2009} and the pioneering work of Paul Slater, cf. Refs. \cite{Slater_1975,Slater_1979}).

In this paper, we explore further the topological architecture of the multi-network of international trade
studying, for the first time, its community structure (see Ref. \cite{fortunato2010} for an overview). 
Detecting the community structure of the ITN and how it correlates with country-specific variables (e.g., size) 
and geography (e.g., distances between countries) is crucial from a international-trade perspective. Indeed,
finding communities in the ITN means identifying clusters of countries that carry tightly interrelated trade linkages among them,
while being relatively less interconnected with countries outside the cluster. To date, only two papers have been trying to
explore the community structure of the ITN \citep{Tzekina2008, Reyes_Comm}. However, they have only studied the aggregate ITN, i.e. the network obtained from total import/export relations between countries irrespective of the specific commodity traded. By focusing on the aggregate ITN only, one indeed neglects the fact that countries actually trade different lines of products and mostly employ imported goods either as inputs to the production process, or as consumption goods. Therefore, identifying clusters of countries from a multi-network perspective may be relevant to better understand what are the countries in the world that tend to trade the same group of products over time and, in turn, uncovering some stylized facts about the actual input-output and supply-demand interdependencies between countries. This may be relevant to predict, for example, to what extent a negative shock hitting a particular industry in a certain region of the world (or in a cluster) may spread and affect the same industry (or closely related ones) in another region of the world (or in another cluster).

Here, we begin addressing this issue by detecting the community structure characterizing the commodity-specific ITN over the period 1992-2003 ($T=12$ years). We employ data about 162 countries and 97 commodities (2-digit disaggregation), to build a sequence of $T$ multi ITNs. We begin by focusing on the 14 top-traded and economically relevant commodities, identifying the community structure of each layer (i.e. groups of countries that mostly trade a given commodity). We then compare commodity-specific community structures with a number of properly-specified community benchmarks. These benchmarks are the community structures obtained from: (i) the aggregate ITN; (ii) the network of geographical closeness (i.e. the inverse of geographical distance) between the 162 countries; (iii) trade-related partitions obtained by detecting the community structure of the regional trade agreement (RTA) network. The main question we ask is whether (and how) commodity-specific community structures are similar to, or differ from, those detected in the benchmark networks. Comparisons are made using the normalized mutual information index (NMI), which is a measure of how close two partitions of the same set of $N$ units are \cite{danon2005}. Understanding whether community structures detected at the commodity-specific level are similar to --- or different from --- those detected in the benchmark networks can shed further light on the topological architecture of the ITN. For example, comparing aggregate and commodity-specific community structures may tell us whether the community structure that we observe at the aggregate trade level can be explained by the aggregation of heterogeneous community structures or, conversely, trade community formation is not affected too much by the type of commodity traded. Similarly, comparing trade-induced communities with those obtained by the network of geographical closeness may help us to understand the extent to which the formation of trade communities is related to geographical distance (as a proxy of trade resistance factors, e.g. trade fees).

The rest of the paper is organized as follows. Section \ref{Sec:Data} describes the databases that we employ in our exercises. Section \ref{Sec:Comms} explains the community detection method that we use in this work. Section \ref{Sec:Results} discusses our main results. Concluding remarks are in Section \ref{Sec:Conclusions}.

\section{Data and Definitions} \label{Sec:Data}

We employ bilateral trade flows data from the United Nations Commodity
Trade Database (UN-COMTRADE; see \texttt{http://comtrade.un.org/}). We build a
balanced panel of $N=162$ countries for which we have commodity-specific
imports and exports flows from 1992 to 2003 ($T=12$ years) in current U.S.
dollars. Trade flows are reported for $C=97$ (2-digit) different commodities,
classified according to the Harmonized System 1996 (HS1996; \texttt{http://www.wcoomd.org/}) \footnote{The 
choice of a 2-digit breakdown of data may be considered insufficient to 
clearly identify homogeneous product lines, but it has been made because in the HS
classification system there is not a unique way to further disaggregate flows by commodities
at a higher number of digits. Notice, however, that network analyses often face a trade off
between the need for a finer disaggregation and the very possibility to obtain connected graphs: 
typically, as soon as 3 or 4 digit data are considered, the resulting graphs easily become not connected,
with the size of the largest connected component quickly decreasing.}.

We employ the database to build a time sequence of weighted, directed
multi-networks of trade where the $N$ nodes are world countries and directed
links represent the value of exports of a given commodity in each year
$t=1992,\dots,2003$ \footnote{Since, as always
happens in trade data, exports from country $i$ to country $j$ are reported
twice (according to the reporting country --- importer or exporter) and
sometimes the two figures do not match, we follow Ref. \citep{Feenstra2005data}
and only employ import flows. For the sake of exposition, however, we follow
the flow of goods and we treat imports from $j$ to $i$ as exports from $i$ to
$j$.}.  As a result, we have a time sequence of $T$ multi-networks
of international trade, each characterized by $C$ layers (or links of $C$
different colors). Each layer $c=1,\dots,C$ represents exports between
countries for commodity $c$ and can be characterized by a $N\times N$ weight
matrix $X_t^c$. Its generic entry $x_{ij,t}^c$ corresponds to the value of
exports of commodity $c$ from country $i$ to country $j$ in year $t$. We
consider directed networks, therefore in general $x_{ij,t}^c\neq x_{ji,t}^c$.
The aggregate weighted, directed ITN is obtained by simply summing up all
commodity-specific layers. The entries of its weight matrices $X_t$ reads:
\beq
x_{ij,t}= \sum_{c=1}^{C}{x_{ij,t}^c},\quad t=1992,\dots ,2003.
\label{eq:aggregate}
\eeq
For the sake of exposition, we shall focus on the most important commodity
networks. Table \ref{Tab:Top14} shows the ten most-traded commodities in 2003,
ranked according to the total value of trade. Notice that they account,
together, for 56\% of total world trade and that the 10 most-traded commodities
feature also the highest values of trade-value per link (i.e. ratio between
total trade and total number of links in the commodity-specific network).
In addition to these 10 trade-relevant commodities, we shall also focus on
other 4 classes (cereals, cotton, coffee/tea and arms), which are less traded
but more relevant in economics terms. The 14 commodities considered account
together for 57\% of world trade in 2003 \footnote{We refer the reader to Ref. \cite{Barigozzi_etal2010pre}
for a thorough analysis of the topological properties of this database from a multi-network perspective.}.

We also employ data about regional trade agreements (RTAs) between world countries taken from the World Trade Organization (WTO) website \footnote{See \texttt{http://www.wto.org/}.}. We build a weighted undirected network with weight matrix $M_t=\{m_{ij,t}\}$ where nodes are countries and a link is weighted according to the number $m_{ij,t}$ of RTAs -- free, multilateral and/or bilateral -- in place between the two countries $i$ and $j$ at year $t$ \citep[cf. also][]{Reyes_Comm}. This sequence of networks may be interpreted as an indicator of how intense are trade agreements between countries over time, i.e. how close countries are in the RTA space. It is well-known from the empirical literature on trade that RTAs are an important determinant of trade flows \cite{Fagiolo2010Jeic}. 

Finally, we build a geographically-related weighted undirected network with weights $s_{ij}=d_{ij}^{-1}$, where $d_{ij}$ are the geographical distances between the most populated cities of country $i$ and country $j$\footnote{Results are robust to alternative distance measures. Data are available at the URL: \texttt{http://www.cepii.fr/}.}. We employ the resulting matrix $S=\{s_{ij}\}$ as a weighted undirected network of geographical closeness between countries, i.e. as the network conveying information on how close countries are in the geographical space. Notice that, traditionally, geographical distance between countries is interpreted as a proxy of all factors that impose some resistance to free trade (transport costs, fees, etc.). 

\section{Community Detection and Comparison}\label{Sec:Comms}
It has been observed that many real networks exhibit a concentration of links
within a special groups of nodes called communities (or clusters or modules).
Such a structural property of a network has also been linked to the presence of 
sub-modules whose nodes have some functional property in common. Therefore,
the detection of the community structure of a given network could help to
discover some hidden feature of its topological architecture.

Despite the intuitive concept of community, a precise definition of 
what a community is represents a challenging issue (see Ref. \cite{fortunato2010}).
In this paper we adopt the well known formulation given in
\cite{newman_girvan2004}: a subgraph is a community if the
number of links (or, more generally, the intensity of interactions)
among nodes in the subgraph is higher than what would be 
expected in an equivalent network with links (and intensities) placed at random. 
This definition implies the choice of a so--called ``null model'', 
i.e. a model of network to which any other network can be compared in 
order to assert the existence of any degree of modularity. 
The most used null model is a random network with the same number
of nodes, the same number of links and the same degree distribution as in the
original network, but with links among nodes randomly placed. 
Based on these concepts, a function called modularity that gives
a measure of the quality of a given network partition into communities has been
introduced in Ref. \cite{newman_girvan2004}. 
The modularity function has been further extended in Ref. \cite{arenas2007} to the
case of weighted directed networks as reported in the following:
\begin{equation}
	\label{eq:mod}
Q = \frac{1}{W}\sum_{ij}\left[ w_{ij} - \frac{w_i^{in}w_j^{out}}{W}\right]\delta_{c_i,cj}
\end{equation} 
where $w_{ij}$ is the weight of the link between $i$ and $j$,
$w_i^{out} = \sum_j w_{ij}$ and $w_j^{in} = \sum_i w_{ij}$ 
are respectively the output and input strengths of nodes $i$ and $j$,
$W = \sum_i\sum_j w_{ij}$ is the total strengths of the network and
$\delta_{c_i,c_j}$ is 1 if nodes $i$ and $j$ are in the same community and 0
otherwise.

In this paper communities are uncovered by optimising the modularity function 
in equation~(\ref{eq:mod}). The optimisation of $Q$ is performed by using a tabu search algorithm
\cite{glover1998}. We shall go back to some critical remarks on the use of modularity-based
community-detection algorithms in the concluding Section.

As discussed in Section \ref{Sec:Intro}, one of the contribution of this paper is
to compare commodity--specific community structures with a proper number 
of community benchmarks (as detailed in the next Section).
To compare community partitions we use the {\em normalised mutual information}
(NMI) measure, as introduced in~\cite{danon2005}. 
To define the NMI index, we need to introduce the confusion matrix. 
Given two community partitions 
$\mathcal{P}_A$ and $\mathcal{P}_B$, the confusion matrix $\mathcal{N}$ 
is defined as a matrix whose
$N_{ij}$-th element is the number of nodes in the community $i$ of the partition
$\mathcal{P}_A$ that appear in the community $j$ of the partition
$\mathcal{P}_B$. The NMI is defined as:
\begin{equation}
\label{eq:nmi}
\mbox{NMI}(\mathcal{P}_A,\mathcal{P}_B) =
\frac{-2\displaystyle\sum_{i=1}^{C_A}\sum_{j=1}^{C_B}N_{ij}log\left(\frac{N_{ij}N}{N_{i.}N_{.j}}\right)}{\displaystyle\sum_{i=1}^{C_A}N_{i.}log\left(\frac{N_{i.}}{N}\right)+\displaystyle\sum_{j=1}^{C_B}N_{.j}log\left(\frac{N_{.j}}{N}\right)}
\end{equation}
where $C_A$ and $C_B$ are respectively the number of communities in
$\mathcal{P}_A$ and $\mathcal{P}_B$, $N_{i.} = \sum_j N_{ij}$,
$N_{.j} = \sum_i N_{ij}$ and $N = \sum_i\sum_j N_{ij}$.
The NMI index is equal to 1 if $\mathcal{P}_A$ and $\mathcal{P}_B$ are identical and assumes a value of 0 if the two partitions are independent.

\section{Results}\label{Sec:Results}
\subsection{Detecting the Community Structure of the Multi ITN}
We begin by studying the connectivity of the multi ITN and the size and concentration of its community structures. All results refer to the aggregate ITN and to the 14 commodity-specific layers as defined in Section \ref{Sec:Data}. In Table \ref{Tab:dens} we show the evolution of the density of the aggregate ITN (computed as the ratio between the number of existing links to the number of all possible links, i.e. $N\dot (N-1)$) and the relative density of commodity-specific networks (relative to the density of the aggregate ITN). We observe a monotonic increase in time of the aggregate ITN density, whereas the relative densities remained almost constant over time in each commodity-specific ITN. This implies an increase in the absolute value of commodity-specific densities. We also observe a relatively high heterogeneity of relative densities across commodity networks, which are always and significantly smaller than the aggregate one. This signals that results obtained using the aggregate ITN may be very different from those obtained looking at single commodity-specific networks (see also below).

While the density measures the concentration of trade links in a network, the size of the largest connected component (LCC) measures its overall level of connectivity. In Table \ref{Tab:greatconn}, we report the size of the LCCs of the aggregate and commodity specific ITNs. While the former is always a completely connected network, this is not always the case for disaggregated cases, see for example arms and cereals. Other commodities, including electronics, optics, plastics and coffee, show instead a large connectivity close to that of the aggregate ITN. This means that their contribution to overall connetivity is very strong. Notice also that for all the commodities we observe an increase in time in the size of the LCC, which is clearly a sign of the increase in the degree of integration of world trade. The largest changes in the size of LCC are observed for arms ($c=92$), cereals ($c=10$) and pharmaceutical products ($c=30$).

We now detect the community structure of both aggregate and commodity-specific ITNs by maximizing a weighted-directed version of the modularity function (see Eq. \ref{eq:mod}). We employ the community structure of the 12 yearly aggregate trade networks with weight matrix as in equation (\ref{eq:aggregate}) as a first benchmark, in order to compare commodity-specific clusters with that obtained from the aggregate trade flows.

The number of communities that we identify in each year and network is shown in Table \ref{Tab:numcomp}. To begin, notice how the aggregate ITN typically displays a smaller number of communities than most of commodity-specific networks, meaning that the latter are more fragmented as far as trade clusters are concerned. In addition, we also observe that the number of communities in the aggregate ITN steadily increases over time, whereas this is not the case for most commodity-specific trade networks. In general, it appears that the smaller the size of the LCC, the higher the number of communities one finds. However, if for every ITN we look at the correlation across time between size of LCC and number of communities, results are different. While for some commodities a larger LCC size implies less communities, for others the opposite holds. In the first group we find coffee and tea ($c=9$), pharmaceutical products ($c=30$), precious stones ($c=71$), and electric machinery ($c=84$). In the second group we find all other commodities and the aggregate ITN. This evidence points to the existence of a large degree of heterogeneity in the number of community structures across commodity-specific networks, suggesting that the results obtained in the case of the aggregate ITN hide a lot of variability in the community structure of commodity-specific networks. This result is in line with similar one obtained in Ref. \cite{Barigozzi_etal2010pre}, where it is shown that many properties of the aggregate ITN (e.g., log-normal distributions of link weights and node-specific characteristics like strength and clustering) are the sheer result of aggregating their counterparts across heterogeneous commodity-specific networks.

\begin{figure}[t]
\begin{center}
\includegraphics[width=18cm,keepaspectratio=true]{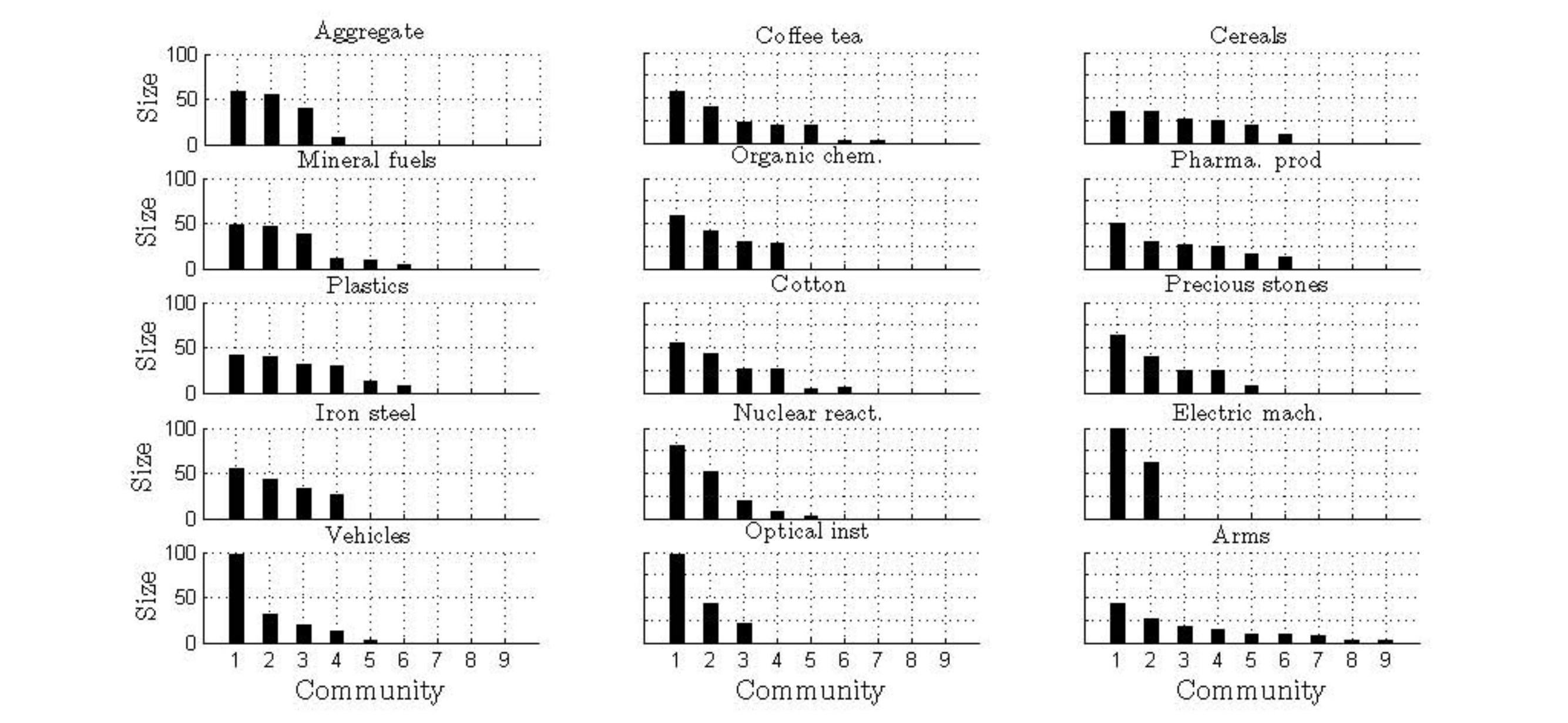}
\caption{{\label{fig:isto}} Cluster-size distributions in 2003.}
\end{center}
\end{figure}

We now turn to a more detailed analysis of the community structure at a commodity-specific level. Figure \ref{fig:isto} shows the distributions of the cluster size in year 2003. Again, the shape of the distributions and their ranges vary a lot across commodities. The commodities that generate the most concentrated community structures are electric machinery ($c=84$), optical instruments ($c=89$), and vehicles ($c=86$), i.e. products that require more scientific knowledge. To draw a more quantitative implication linking products and concentration of cluster-size distributions, we compute the normalized Herfindahl index (H), a synthetic measure of concentration of cluster size distributions. The index H, for a given commodity $c$ and in a given year $t$, is defined as:
\begin{equation}
\mbox{H}_{t}^c=\frac 1 {1-\frac 1 N}\l\{\l[\sum_{i=1}^{n_{X_t^c,t}}\l(\frac{m_{t}^c(i)} N\r)^2\r ]-\frac 1 N \r\},\quad t=1992,\dots ,2003,
\end{equation}
where $n_{X_t^c,t}$ is the number of communities identified in the network $X_t^c$ and $m_{t}^c(i)$ is the number of countries in the $i$-th community in year $t$ for commodity $c$. The index ranges between  $0$ (no concentration at all) and $1$ (maximum concentration). Table \ref{Tab:HI} reports the values of $\mbox{H}_{t}^{c}$ for all networks and time periods. It is easy to notice that for the aggregate ITN there has been a decrease in concentration over time. This may be interpreted as a sign of the globalization process, as this pattern suggests that an increasing number of countries are participating to world trade over time. Indeed, while in 1992 we observe only 2 communities of about 80 countries each (one with Europe, Russia and Africa, the other with America and Asia), in 2003 a new community emerges, driven by China and India. At the commodity-specific level, an increase in H is observed also for coffee and tea ($c=9$), mineral fuels ($c=27$), pharmaceutical products ($c=30$), arms ($c=92$). However, for some other commodities we observe a decrease in H over time, see e.g. organic chemicals ($c=29$), plastics ($c=39$), and cotton ($c=52$). This means that trade for those commodities has become less and less centralized and increasingly occurred among smaller and more dispersed groups of countries.

\subsection{Describing Trade Communities}
\begin{figure}[t]
\begin{center}
{\includegraphics[width=16cm, keepaspectratio=true]{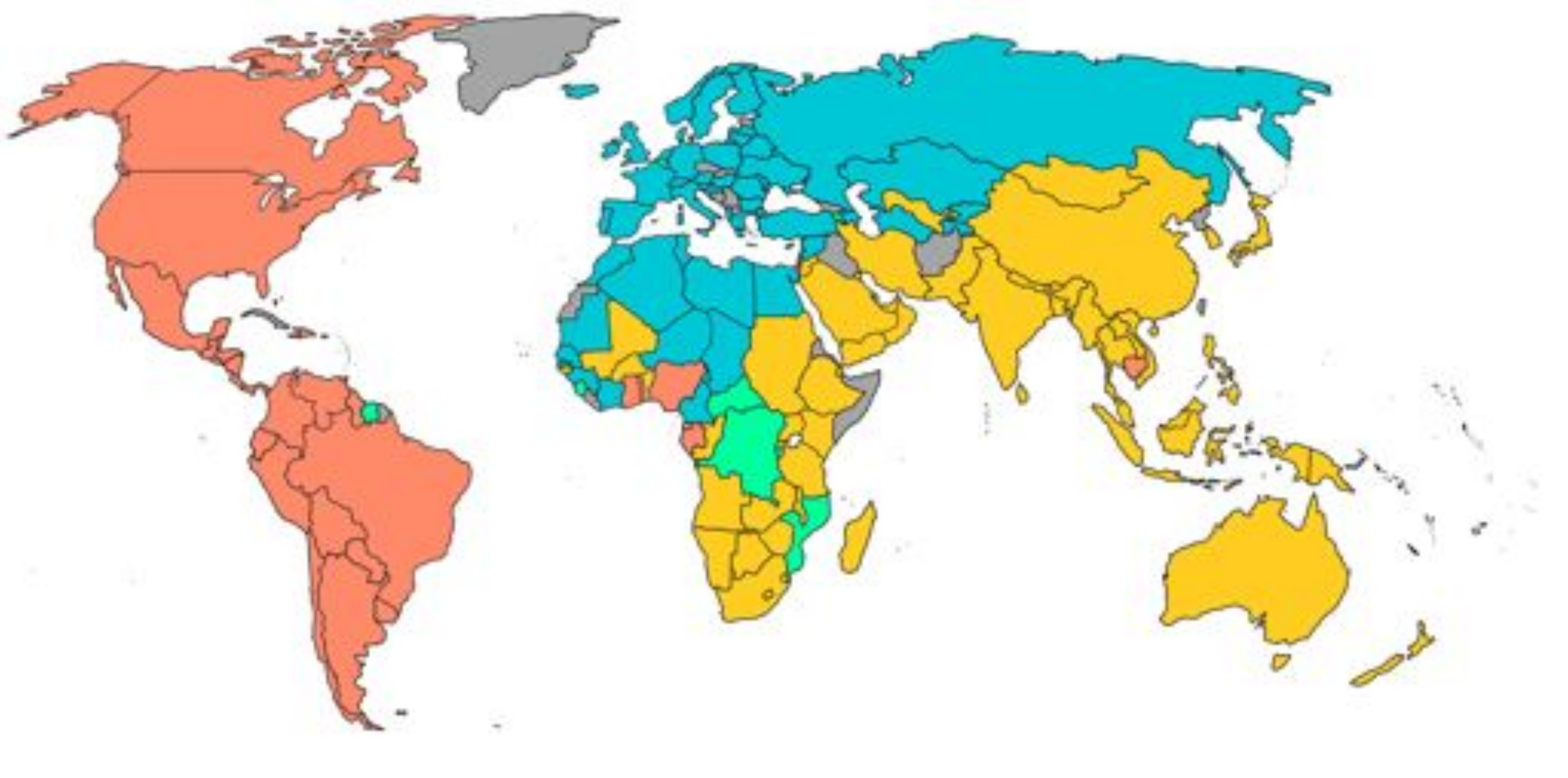}}
\caption{{\label{fig:map_agg}}World map showing communities of aggregate ITN in 2003. In gray countries not belonging to any community or for which no data are available.}
\end{center}
\end{figure}

\begin{figure}[t]
\begin{center}
\subfigure[Coffee and tea $c=9$.]{\includegraphics[width=7cm, keepaspectratio=true]{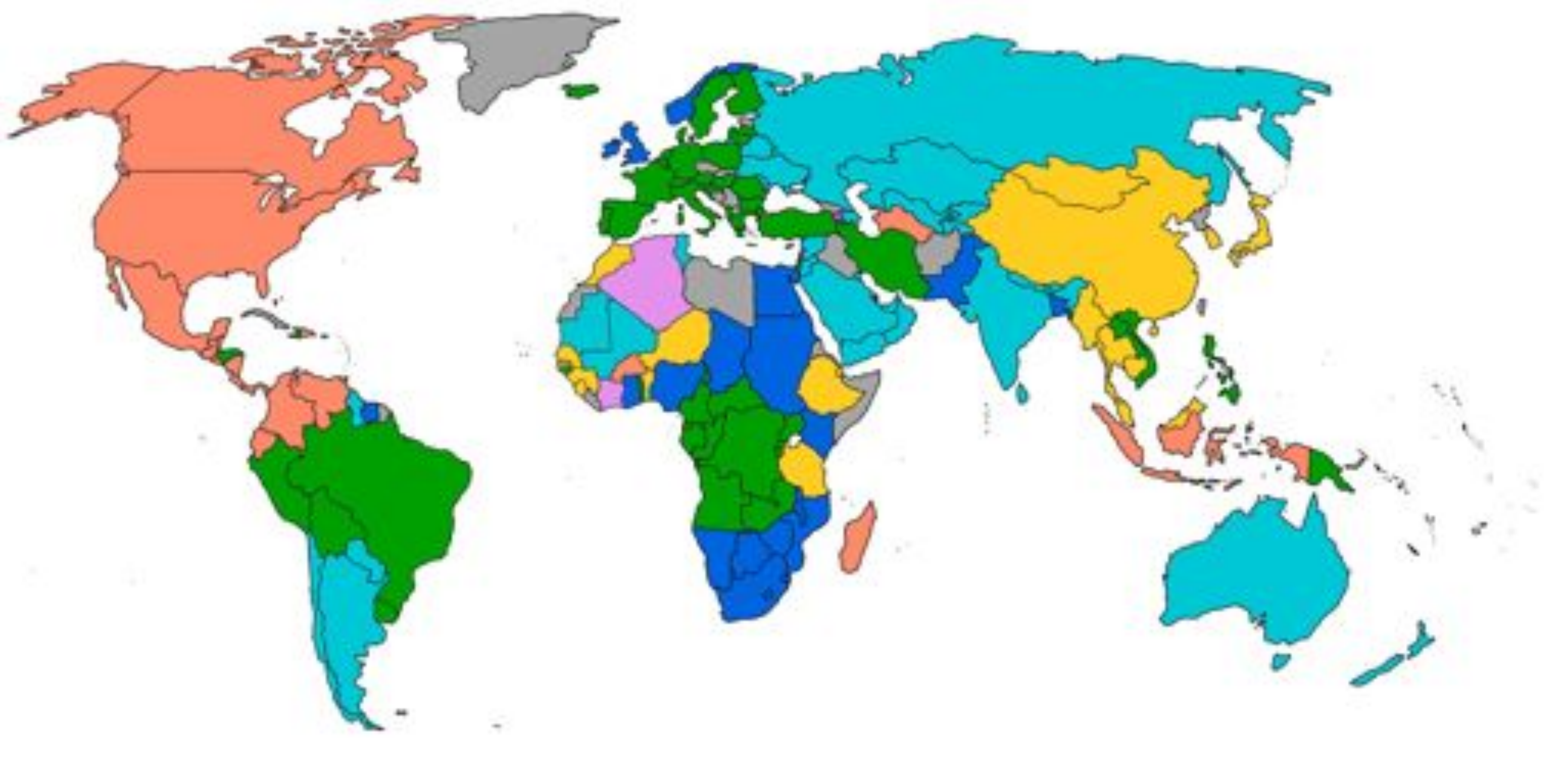}}
\subfigure[Cereals $c=10$.]{\includegraphics[width=7cm, keepaspectratio=true]{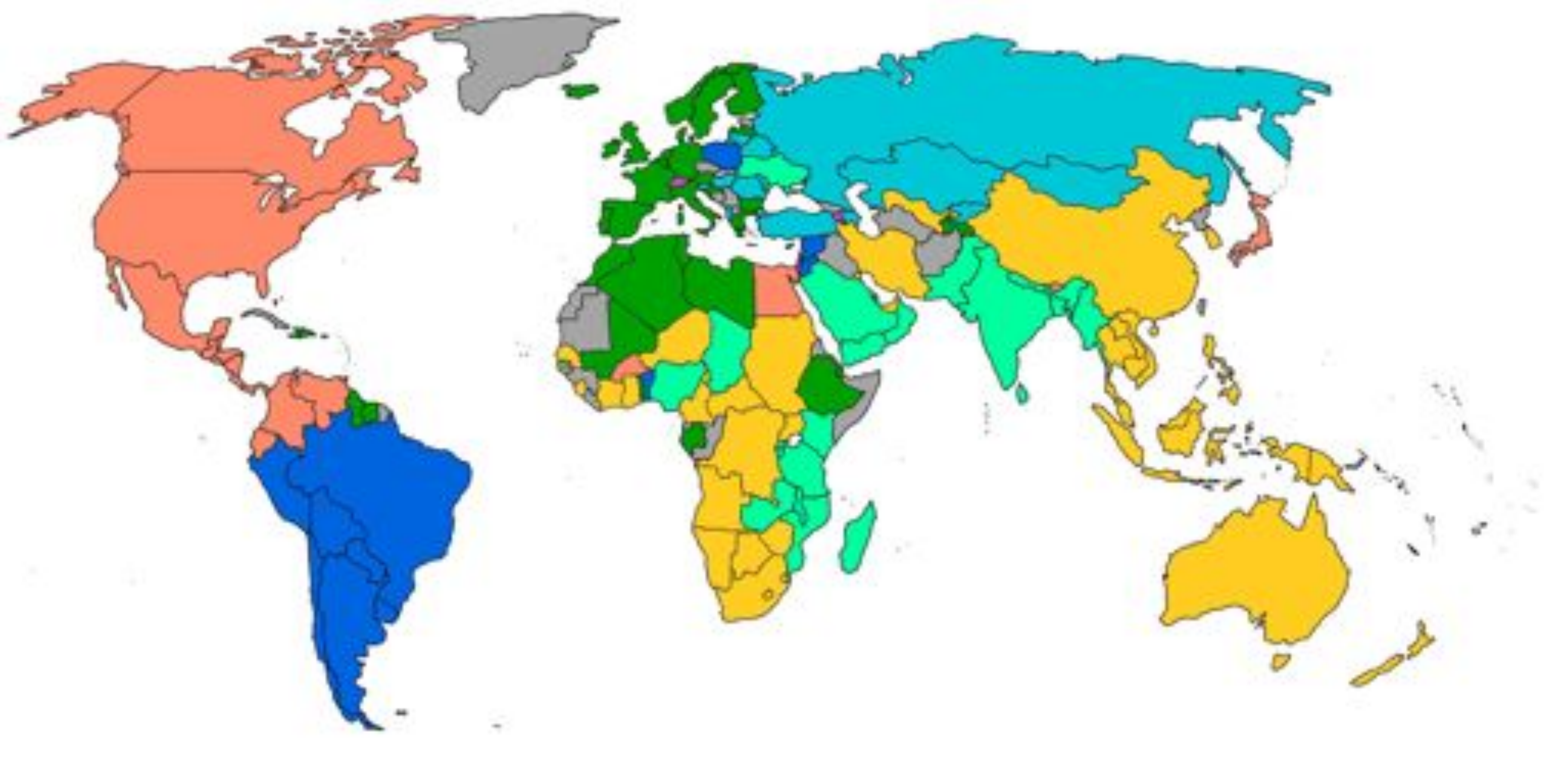}}\\
\subfigure[Mineral fuels $c=27$.]{\includegraphics[width=7cm, keepaspectratio=true]{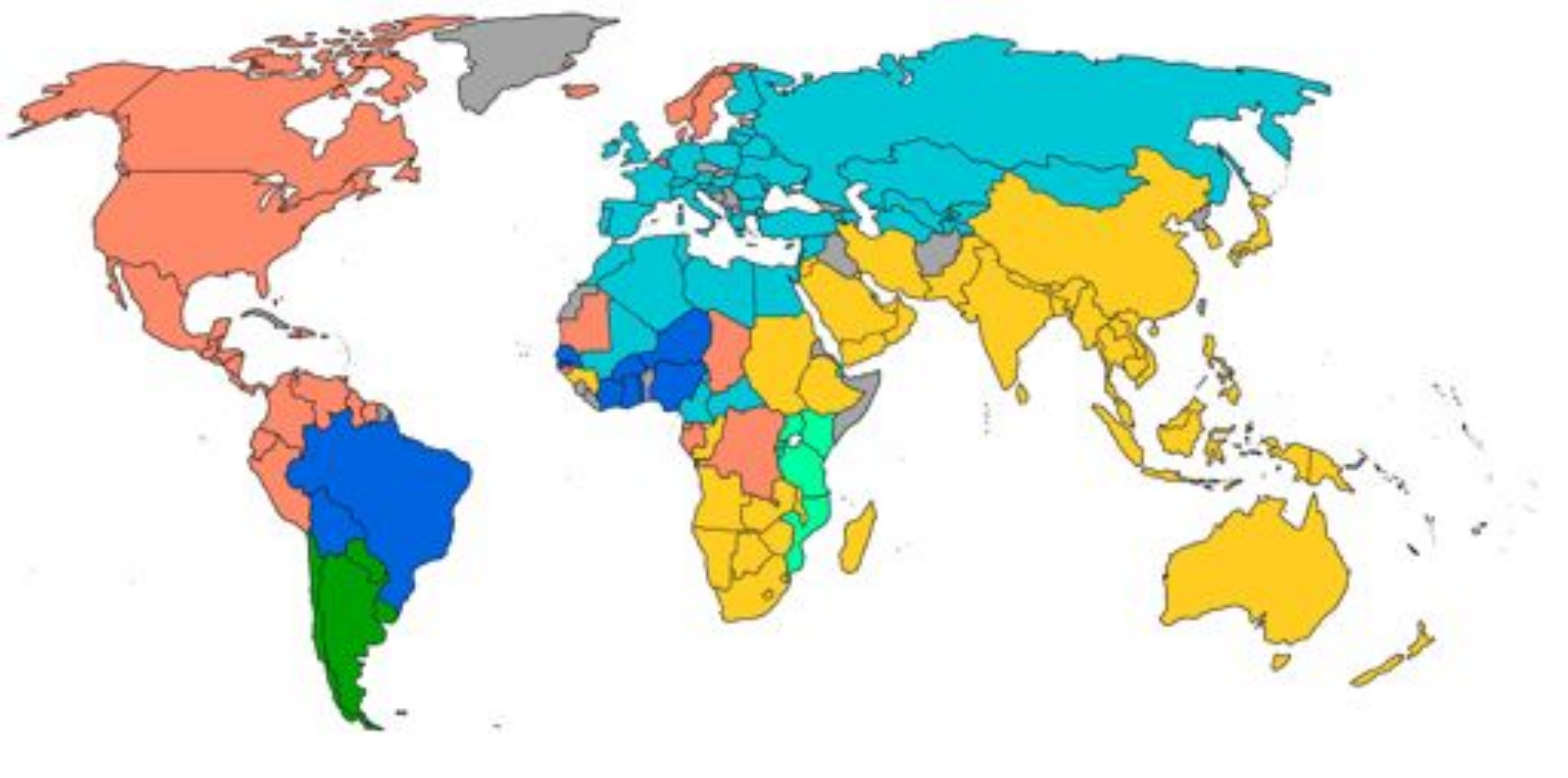}}
\subfigure[Organic chemicals $c=29$.]{\includegraphics[width=7cm, keepaspectratio=true]{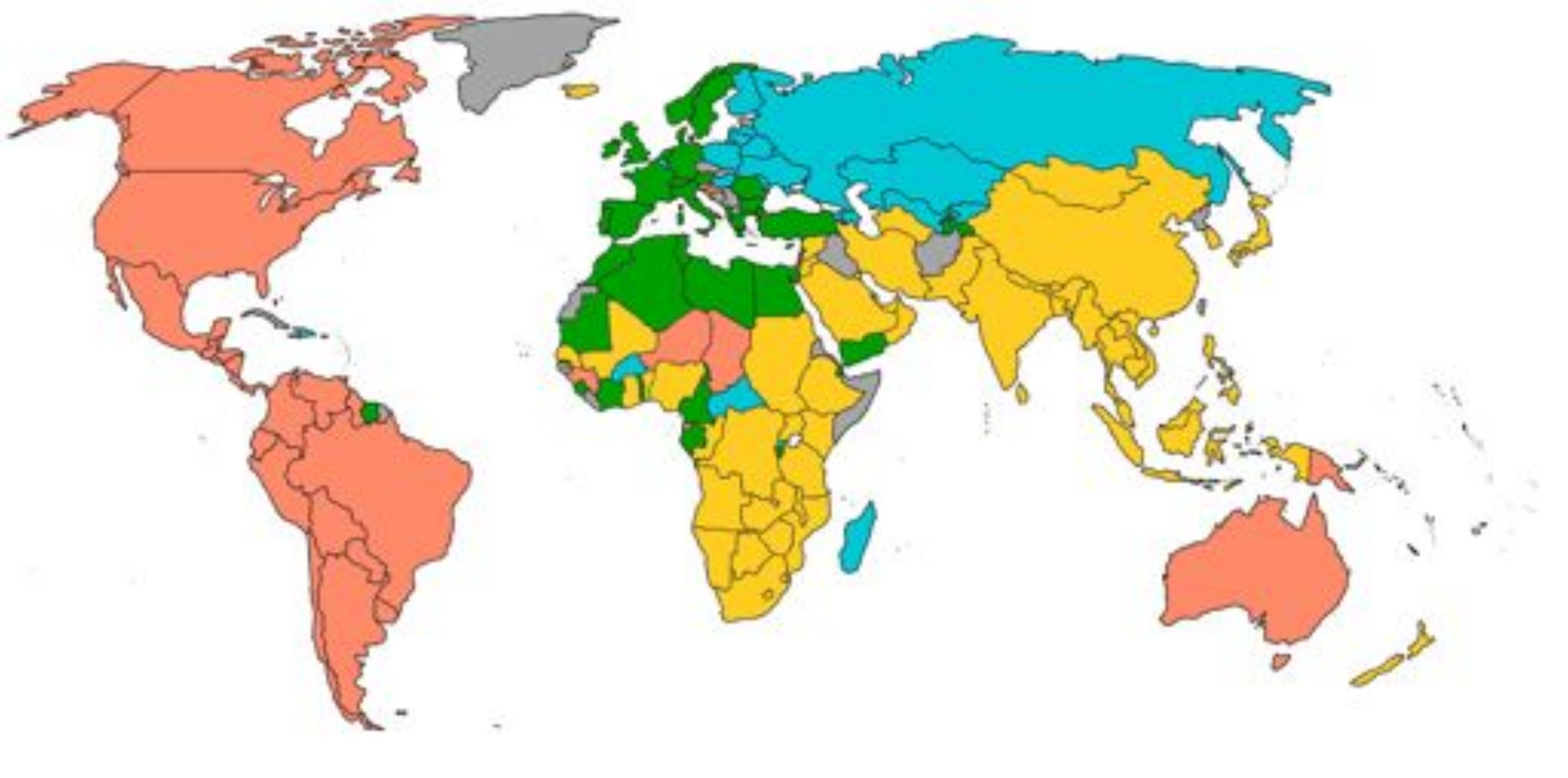}}\\
\subfigure[Pharmaceutical products $c=30$.]{\includegraphics[width=7cm, keepaspectratio=true]{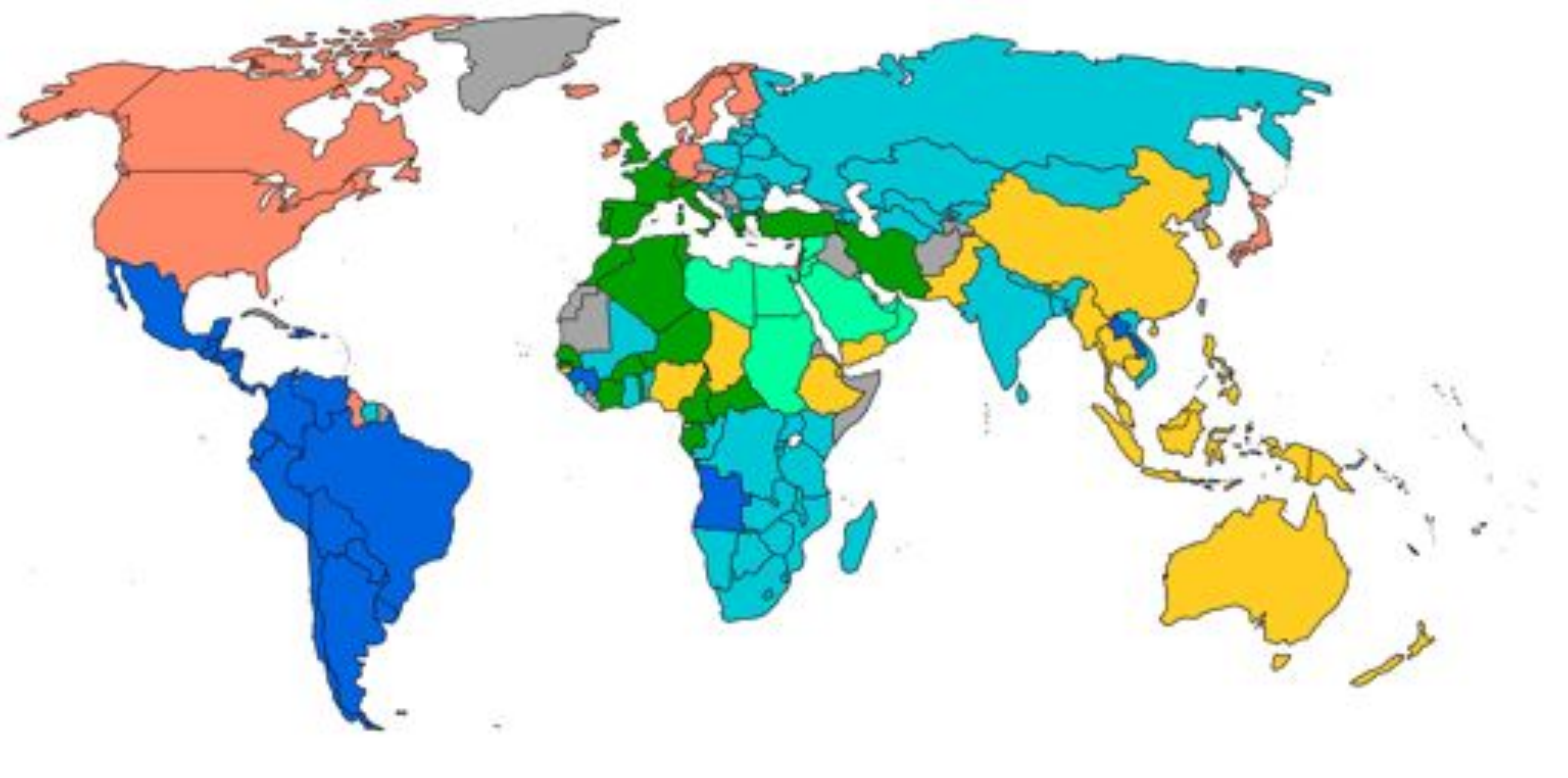}}
\subfigure[Plastics $c=39$.]{\includegraphics[width=7cm, keepaspectratio=true]{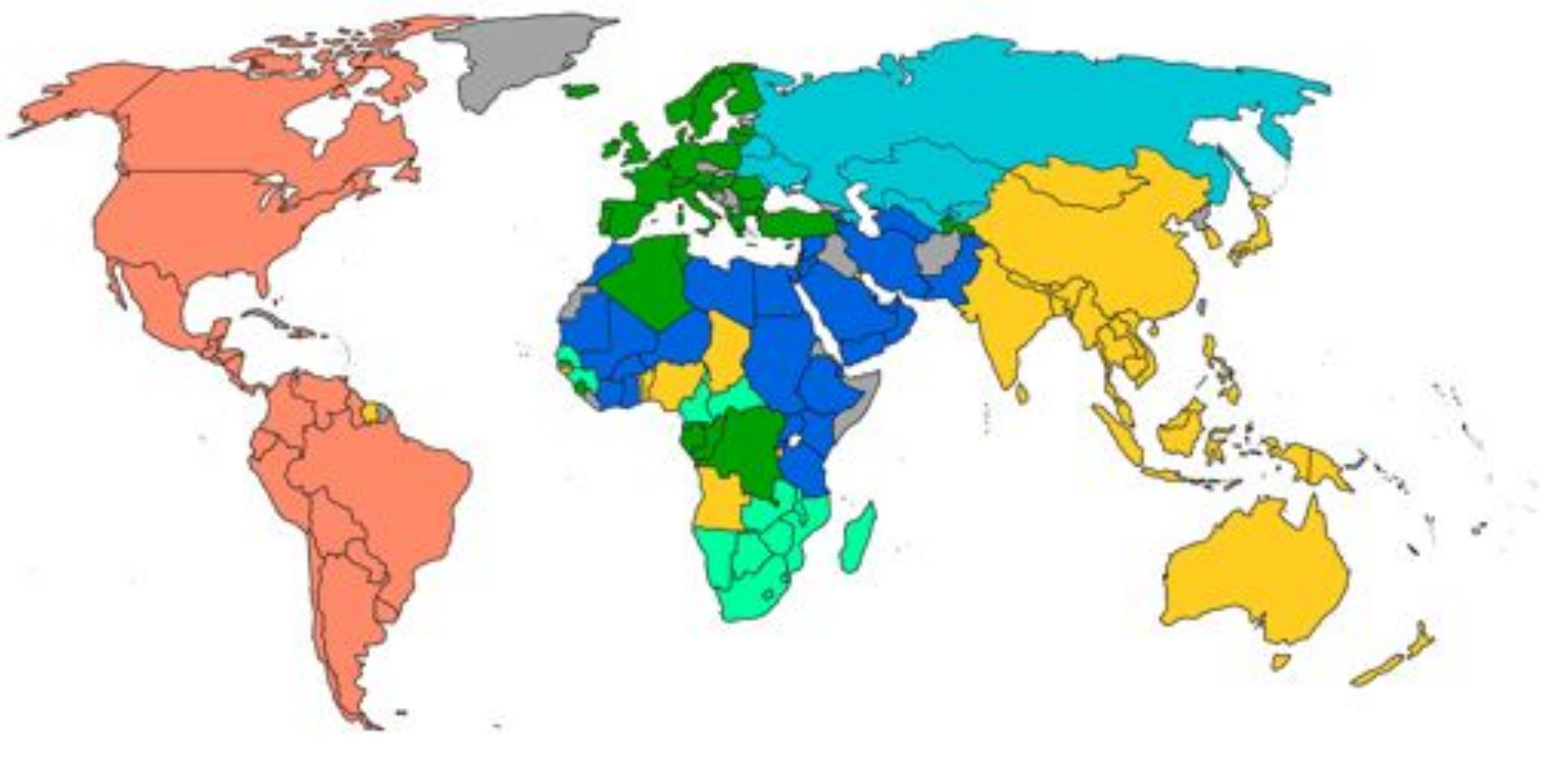}}\\
\subfigure[Cotton $c=52$.]{\includegraphics[width=7cm, keepaspectratio=true]{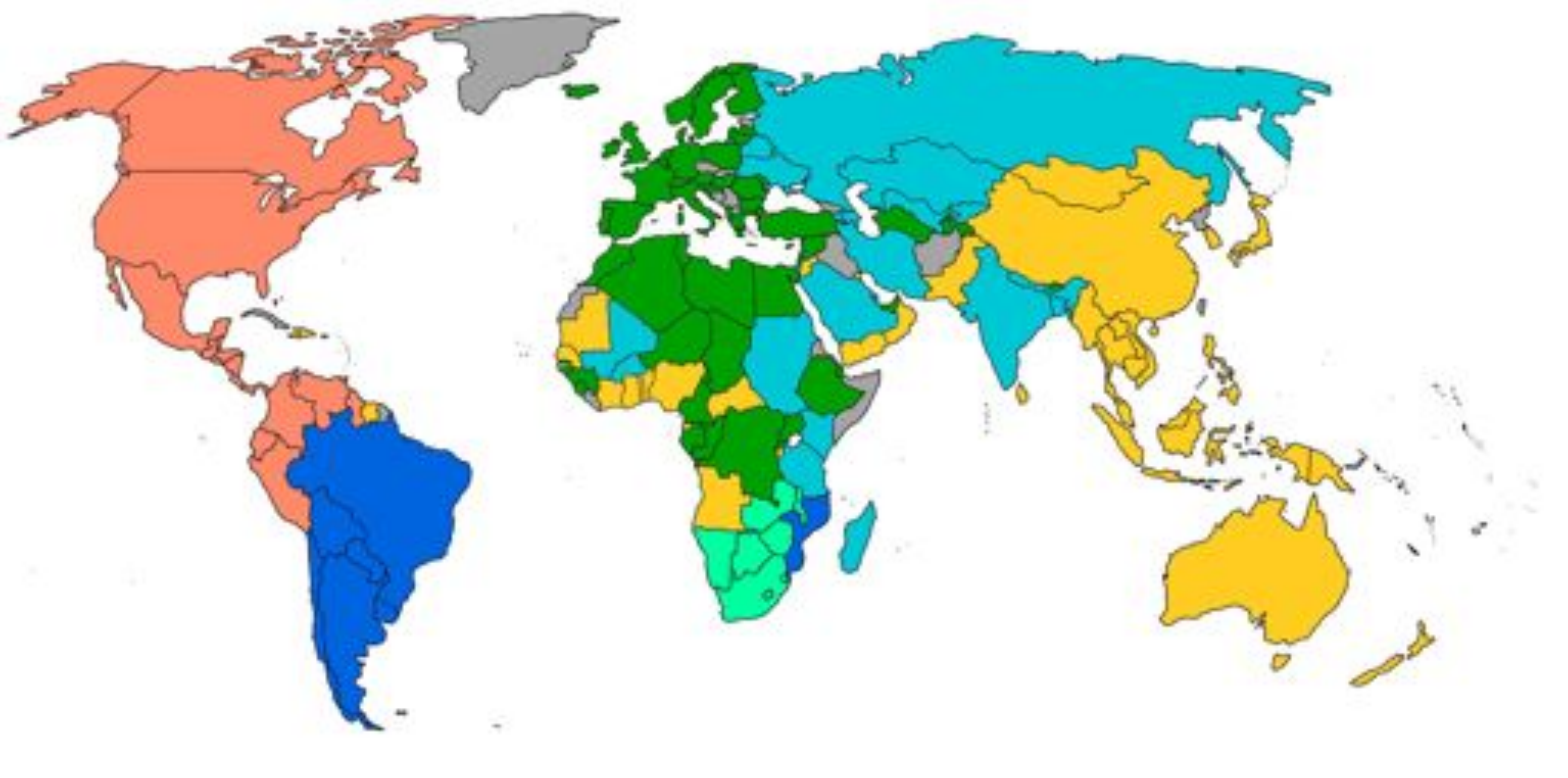}}
\subfigure[Precious stones $c=71$.]{\includegraphics[width=7cm, keepaspectratio=true]{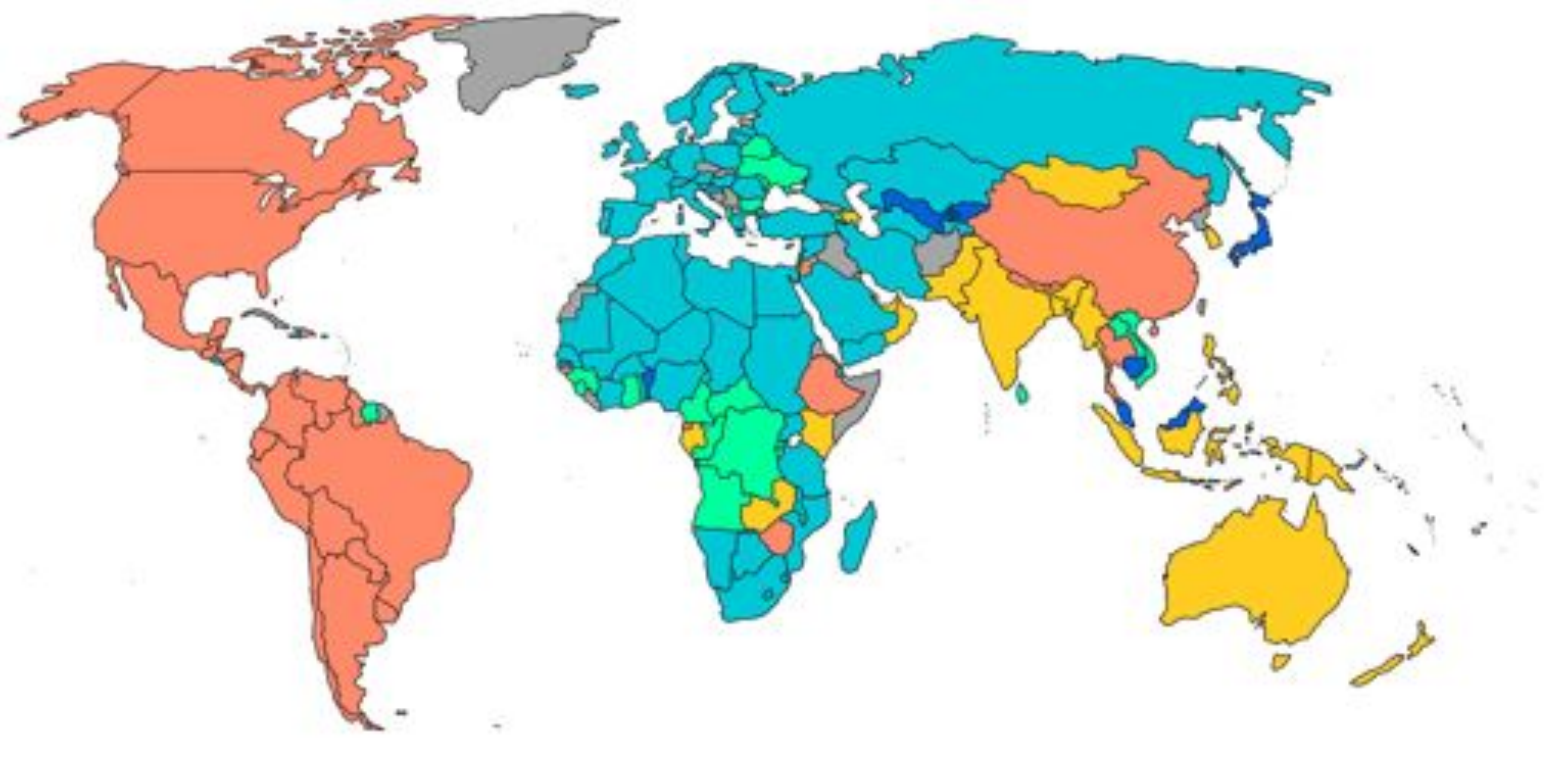}}\\
\subfigure[Iron and steel $c=72$.]{\includegraphics[width=7cm, keepaspectratio=true]{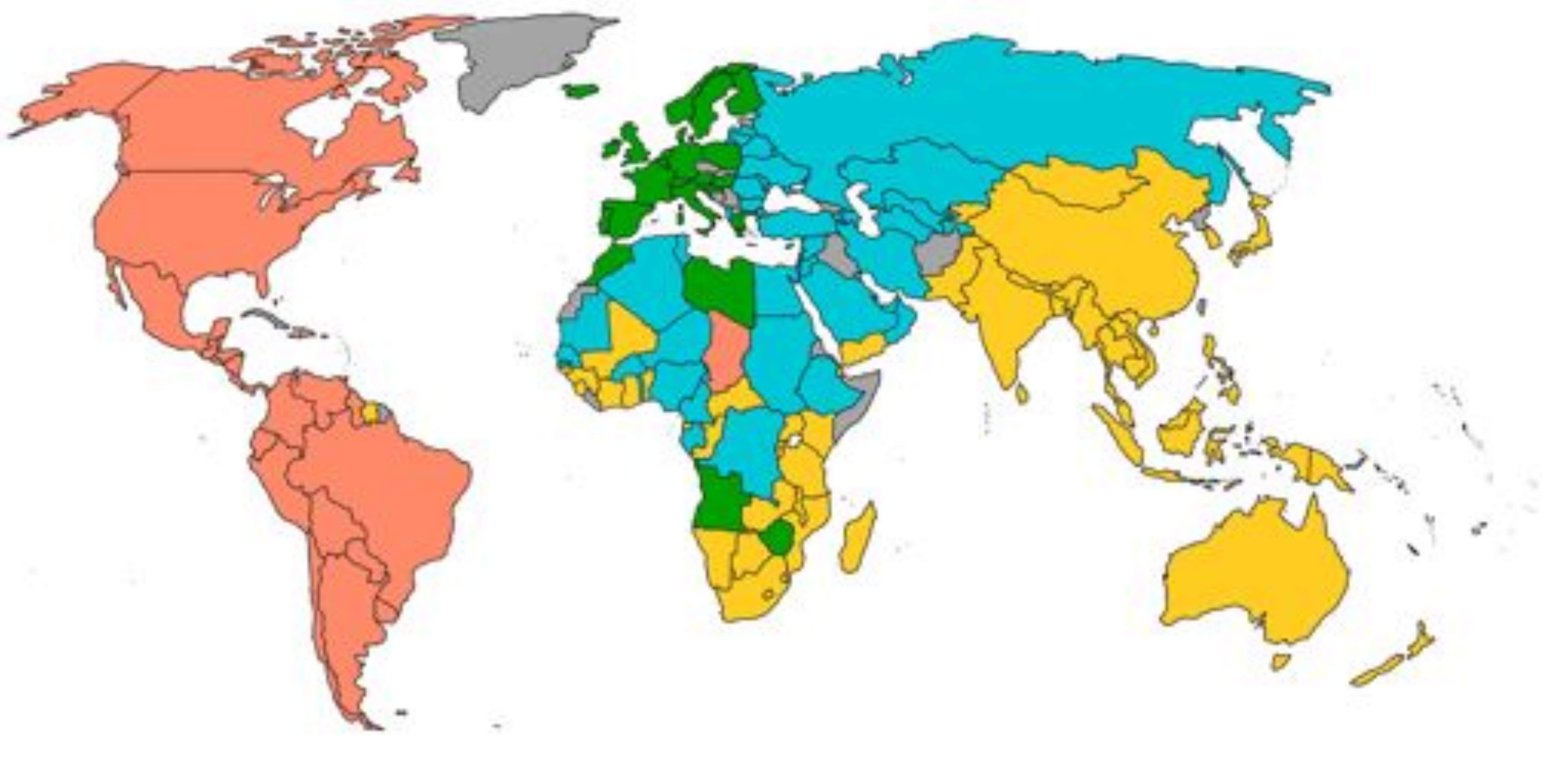}}
\subfigure[Nuclear reactors $c=83$.]{\includegraphics[width=7cm, keepaspectratio=true]{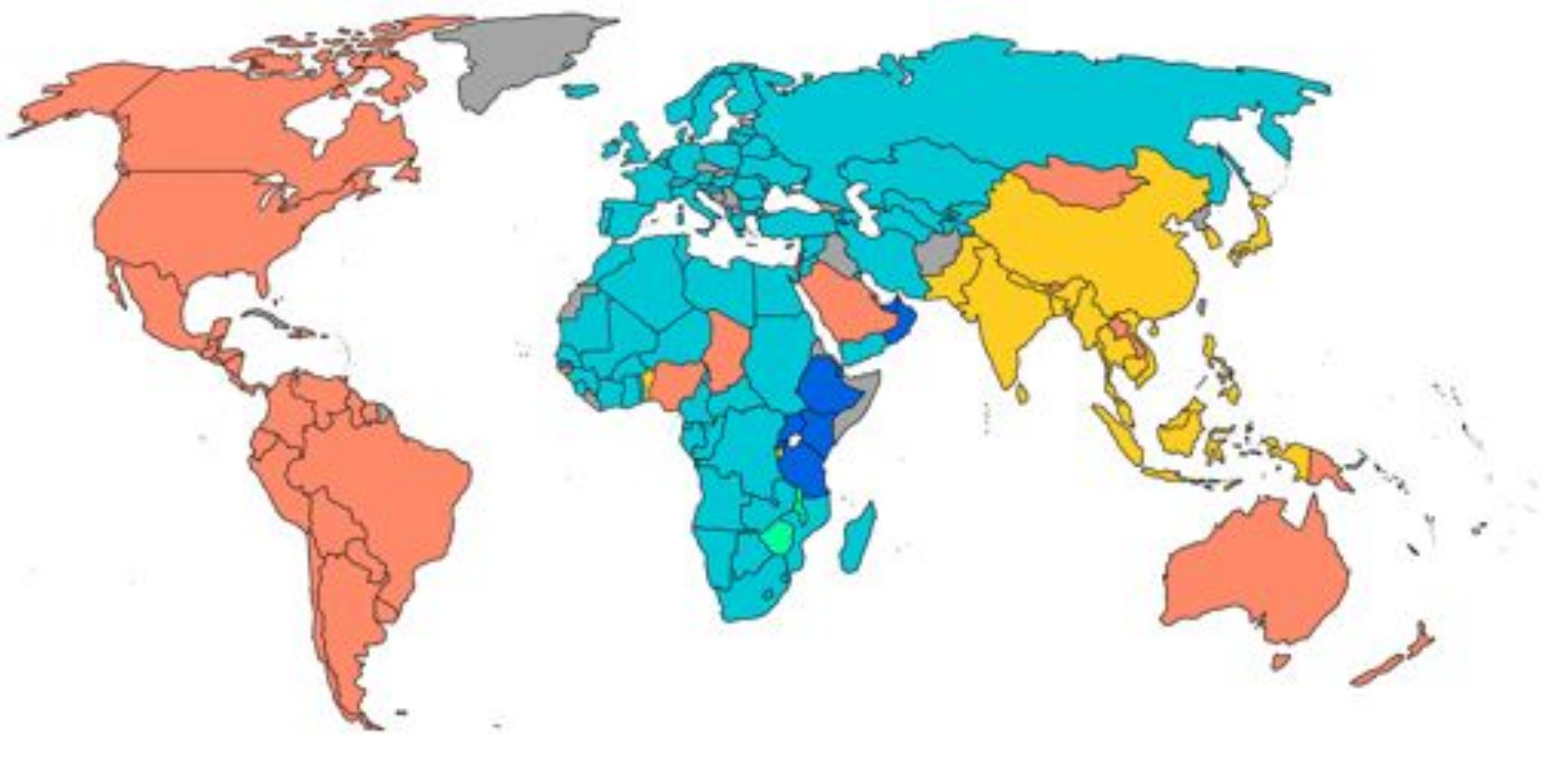}}\\
\caption{{\label{fig:map_disagg}}World maps showing trade communities of commodity specific ITNs in 2003. In gray countries not belonging to any community or for which no data are available.}
\end{center}
\end{figure}

\begin{figure}[t]
\begin{center}
\subfigure[Electric machinery $c=84$.]{\includegraphics[width=7cm, keepaspectratio=true]{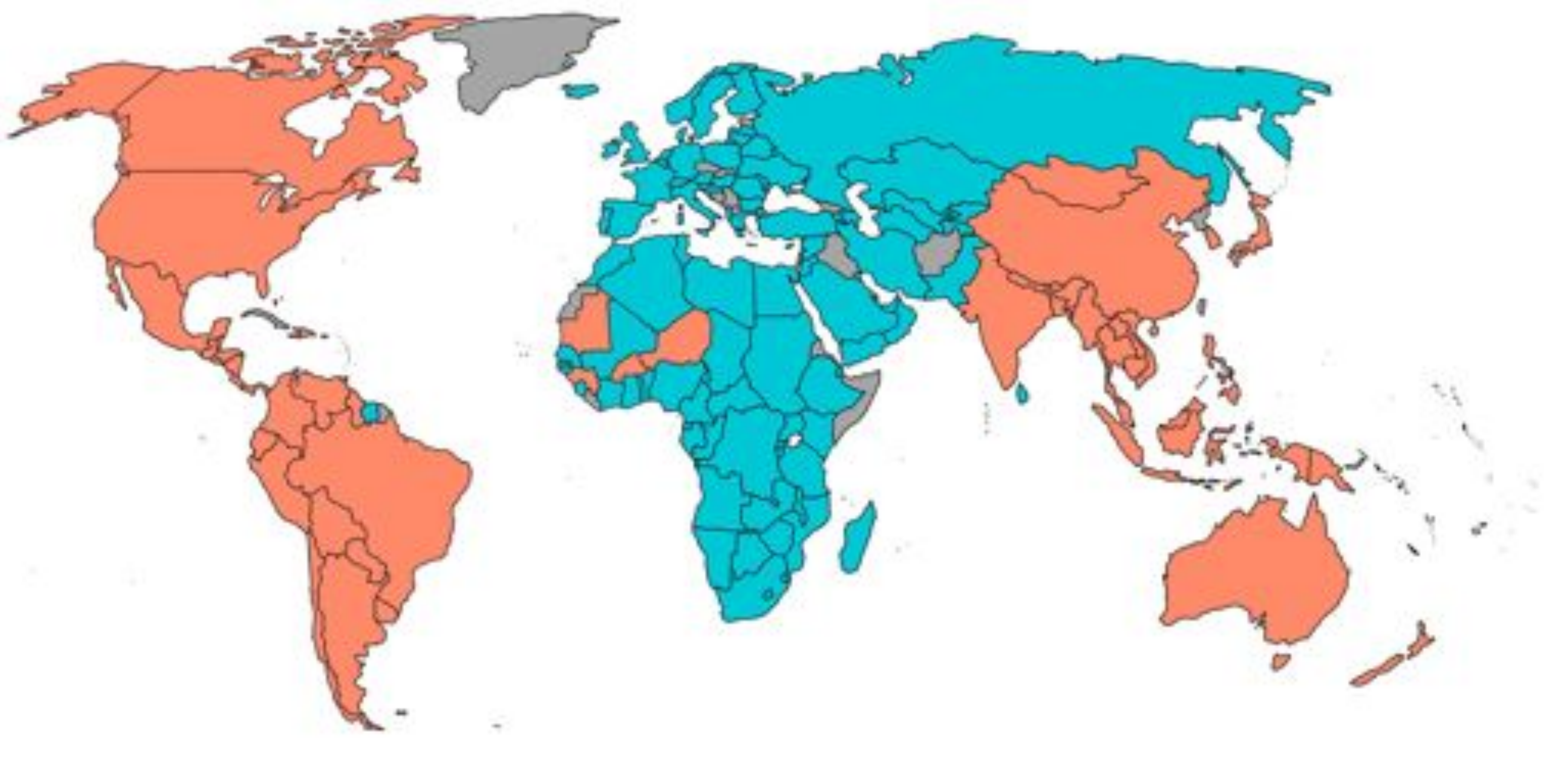}}
\subfigure[Vehicles $c=86$.]{\includegraphics[width=7cm, keepaspectratio=true]{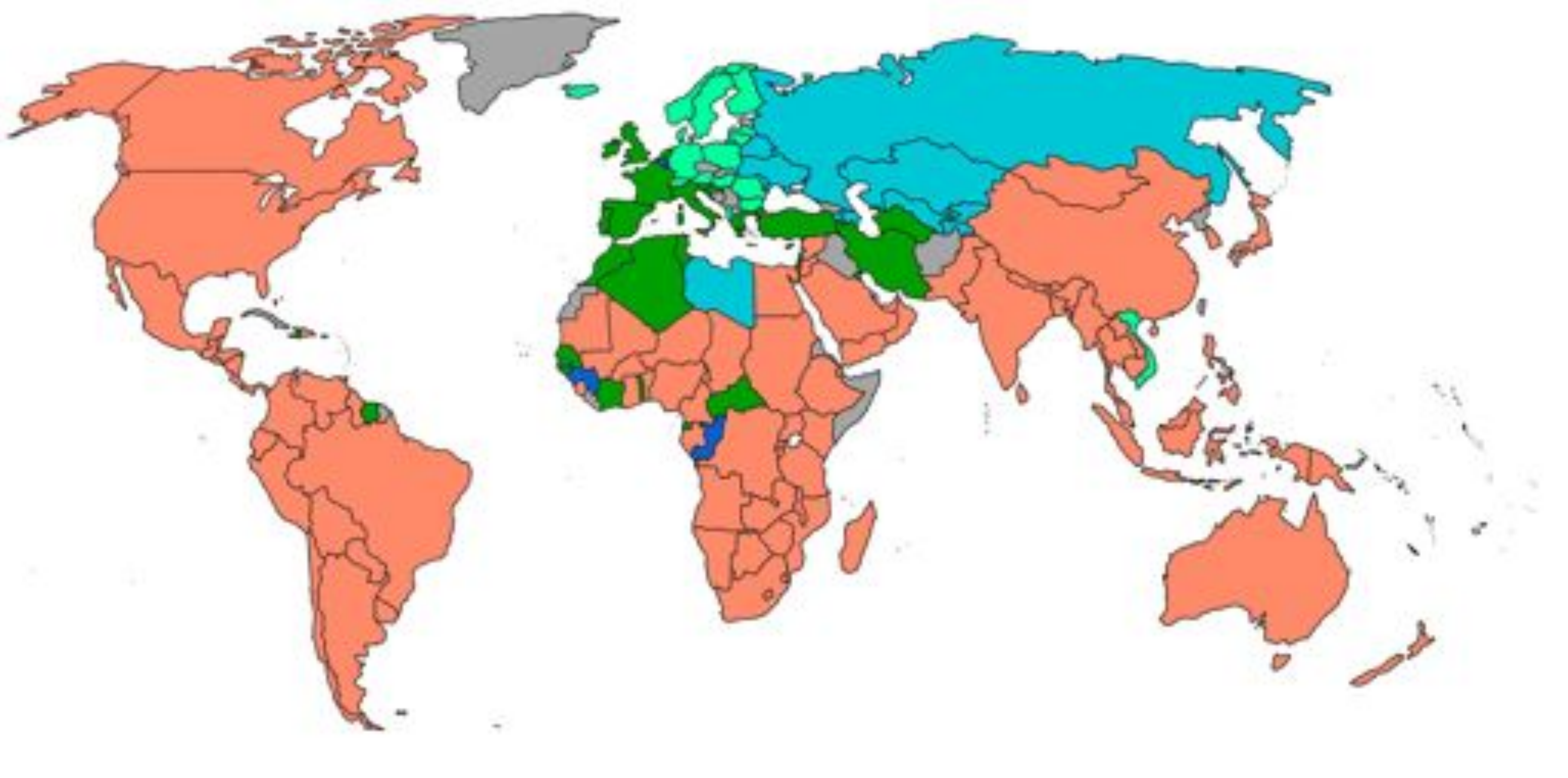}}\\
\subfigure[Optical instruments $c=89$.]{\includegraphics[width=7cm, keepaspectratio=true]{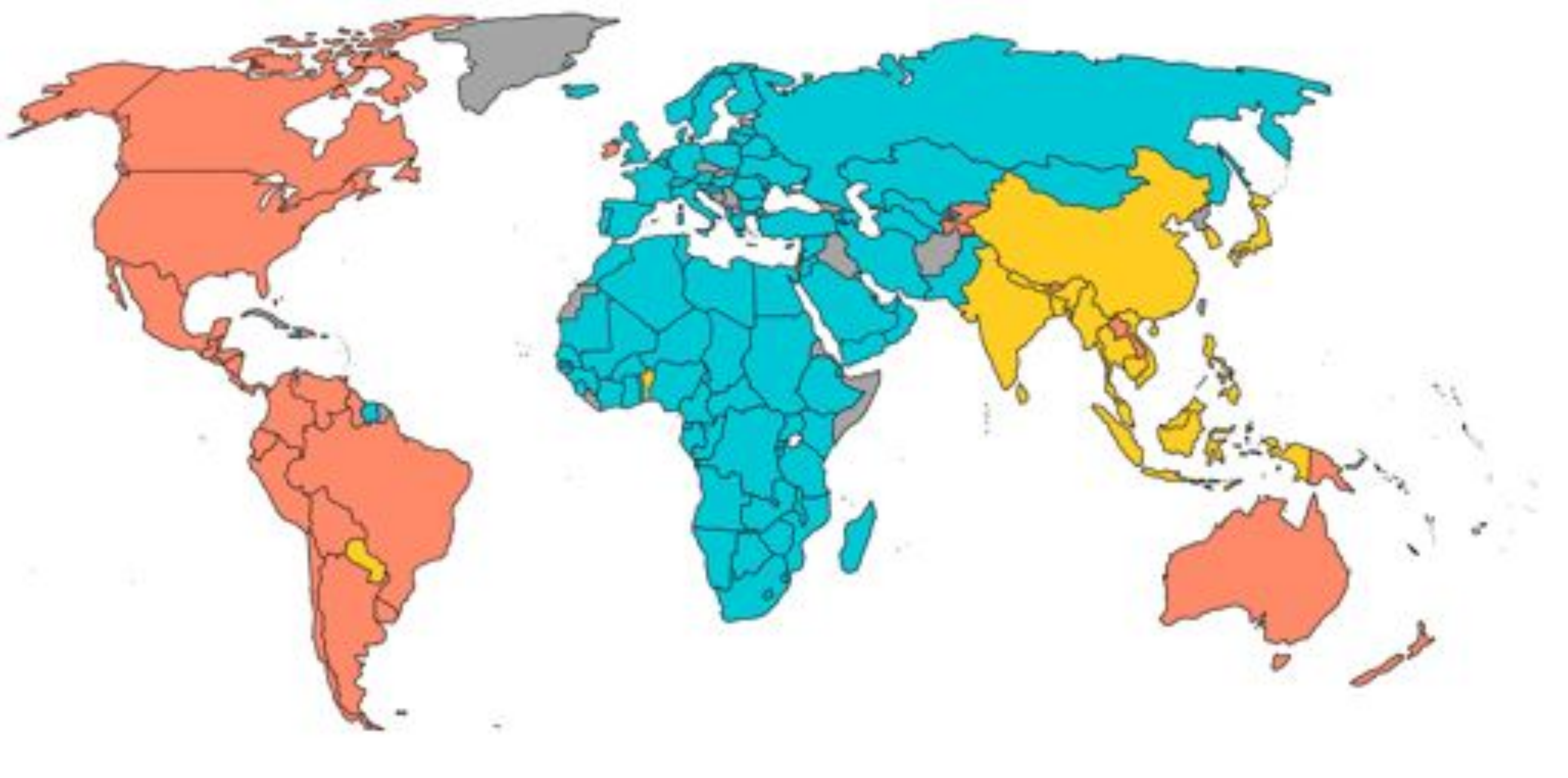}}
\subfigure[Arms $c=92$.]{\includegraphics[width=7cm, keepaspectratio=true]{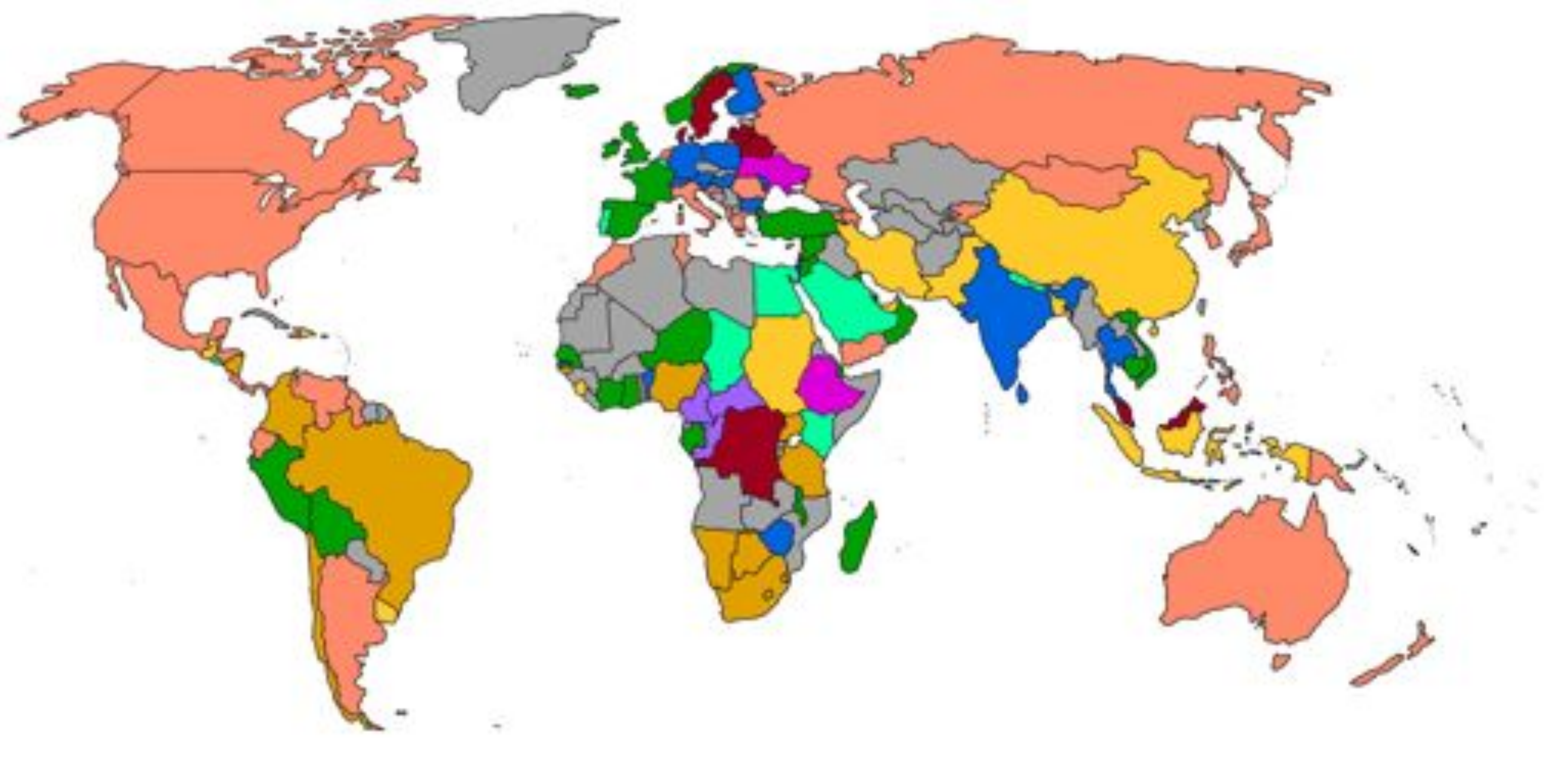}}
\caption{{\label{fig:map_disagg2}}World maps showing trade communities of commodity specific ITNs in 2003. In gray countries not belonging to any community or for which no data are available.}
\end{center}
\end{figure}

A useful way to visually describe community structure in the ITN is to employ colored world maps, where countries belonging to the same communities are associated to the same color. Figures \ref{fig:map_agg}-\ref{fig:map_disagg2} report world maps depicting the community structure detected in 2003, for both the aggregate ITN and for the 14 commodity-specific networks. 

Notice that this visual device also allows us to informally correlate community structures with geographical considerations (we shall go back to a more formal analysis of this issue below). For example, most of the networks studied exhibit the presence of an American cluster composed of US and Canada (and often linked to Latin America), a European cluster (sometimes connected to North Africa), an Asian cluster consisting of China (and in many cases of India, Indochina and Australia) and finally a Russian community (sometimes linked to the European cluster). Africa and Middle East are often split, independently of the commodity examined, among the other groups. This already suggests that geographical (and socio-political) factors are very important to explain the formation of community structure in the ITN. 

Apart from the regularities above, commodity-specific community structures often differ in a relevant way among each other. In what follows, we highlight some economically-relevant features of aggregate and commodity-specific community structures in 2003. We focus on 7 commodity classes, those exhbiting the most economically relevant patterns (the remaining 7 classes did not show such explicit regularities). Due to the relatively strong persistence over time of ITNs topological architecture (see Refs. \cite{Fagiolo2009pre} and \cite{Barigozzi_etal2010pre} for a discussion), similar considerations also hold for other years.

\begin{enumerate}

\item \textbf{Aggregate ITNs}: The world is divided in three major communities which follow a geographical pattern: i) North and Latin America, ii) Europe, Russia, and North Africa, iii) China, India, Japan, Middle East, Australia and Sub-Saharan Africa. Two exceptions concern Africa: Nigeria and Ghana belong to the American community and we observe a minor separated community containing Belgium and the Democratic Republic of Congo, a former Belgian colony.

\item \textbf{Coffee and tea}: We identify two communities containing coffee drinking and producing countries: i) Europe, Brazil, Peru, and Central African countries, ii) North America, Central America, Colombia and Venezuela. We also identify two communities of mainly tea drinking and producing countries: i) United Kingdom, South African and North-East African countries, Pakistan and Bangladesh. Finally, there exist two mixed communities, but probably more connected with tea trade: i) India, Middle East, Russia, Australia, Argentina, Chile, ii) China, Japan, and Indochina. 

\item \textbf{Cereals}: The big producers of cereals belong each to a separate community: i) North America, ii) South America, iii) Russia. China and India are in separate communities, too. Finally, it is interesting to notice that Europe belongs to yet another separate cluster. Despite being not a big producer, but a big consumer, Europe is not an open market for agricultural products. This finding may be linked to the protectionist agricultural policies of the European Community.
 
\item \textbf{Mineral fuels}: China and India have tight links with Middle East, Europe has links with Russia and North Africa, Brazil with Nigeria, and North America with Norway, which is one of the largest oil producers in the world.

\item \textbf{Precious stones}: In this case America and China belong to the same community. Europe, Russia, North and South Africa are the members of the largest community. Interestingly, countries rich of diamonds as Democratic Republic of Congo, Angola, and Sierra Leone, belong to a unique community containing also Israel. Finally, Australia, Indonesia, and India belong to another cluster.

\item \textbf{Electric machinery}: There are only two communities strictly related to geographical distance. Indeed, countries within a community share common borders. One contains America, China, Japan, India and Australia. The other one contains Europe, Russia, Africa and the Middle East.

\item \textbf{Vehicles}: The world market for vehicles has a huge community containing America, China, Japan, India, Australia and almost all African countries. This may reflect the high diffusion of Japanese cars in Africa. Russia is still a closed market containing all former Soviet republics. Finally, Europe is divided in two communities, which have almost no members outside the continent: a finding that seems to reveal a protectionist market for vehicles in Europe.

\item \textbf{Arms}: The community structure for arms is highly fragmented and therefore difficult to interpret. Moreover, many countries seem not to belong to any community. Interestingly, these countries are often those were civil wars or in general social instability are most likely to be (or to have been) present. This is the case of Mozambique, Zambia, Angola, Guinea, Myanmar and Central Asian countries. It is unlikely that these countries do not participate in arms trade, but it is not surprising that our data do not reveal this, as probably that kind of trade relationships are not official. Finally, Africa is the most fragmented continent and almost all communities found contain some African countries.

\end{enumerate}

\subsection{Comparing Community Structures}
In this Section, we explore more quantitatively commodity-specific community structures by using the NMI index introduced in Section \ref{Sec:Comms}. 

To begin with, we ask to what extent community structures are stable over the time interval considered. To do that, we compare the partitions obtained at time $t$ and $t+1$ for $t=1992,\dots,2002$. More precisely, for the aggregate ITN and for any $c$, we compute the quantity $\mbox{NMI}(\mathcal{P}_{t}^c,\mathcal{P}_{t+1}^c)$, where $\mathcal{P}_{t}^{c}$ is the partition of our $N$ countries in year $t$ for commodity $c$. This gives a measure of stability over time of community structures (see Table \ref{Tab:NMI_time}). Notice how the smallest values of NMI (large community structure changes) are observed in the early 1990s. In more recent years, on the contrary, NMIs have been larger, meaning weaker changes in the composition of communities from year to year. If one instead compares partitions in 1992 with those in 2003 (i.e., one computes the quantity $\mbox{NMI}(\mathcal{P}_{1992}^c,\mathcal{P}_{2003}^c)$), it turns out that the stronger changes are associated to coffee and tea ($c=9$), pharmaceutical products ($c=30$), and arms ($c=92$). The most stable community structures are instead those of aggregate trade, plastics ($c=39$), optical instruments ($c=89$), mineral fuels ($c=27$), iron and steel ($c=72$), and cotton ($c=52$). Notice also that, on average, the majority of commodity-specific community structures were less stable than that of the aggregate network. Again, this suggests a strong mismatch between aggregate and disaggregated properties.

We then compare, in each given year, partitions associated to the aggregate ITN with those associated to commodity-specific networks by computing the quantities $\mbox{NMI}(\mathcal{P}_{t}^{all},\mathcal{P}_{t}^c)$, where $\mathcal{P}_{t}^{all}$ is the partition obtained from the aggregate ITN. This exercise is meant to ask more quantitatively the question whether the aggregate community structure can well predict those obtained at the commodity-specific level, or, put it differently, the extent to which the community structure of any commodity-specific network contributes to shape (or is able to predict) that observed at the aggregate-trade level. Inspection of Table \ref{Tab:NMI_agg_disagg} shows that NMI values are increasing in time for almost all commodities. This means that commodity-specific community structures are becoming more and more similar to the aggregate one, i.e. that the role of all commodities in shaping the aggregated community structure has increased in time. In particular, we observe the largest increase in NMI from 1992 to 2003 for mineral fuels ($c=27$), plastics ($c=39$), iron and steel ($c=72$), pharmaceutical products ($c=30$), and organic chemicals ($c=29$). Moreover, for all the years considered, mineral fuels and plastics appear to be the commodities whose community structure is the most similar to the aggregated one. Overall, a major role for the chemical sector emerges from these results, in that they are the ones whose country partition better mimics the aggregate one.

\begin{figure}[t]
\begin{center}
\includegraphics[width=10cm,keepaspectratio=true]{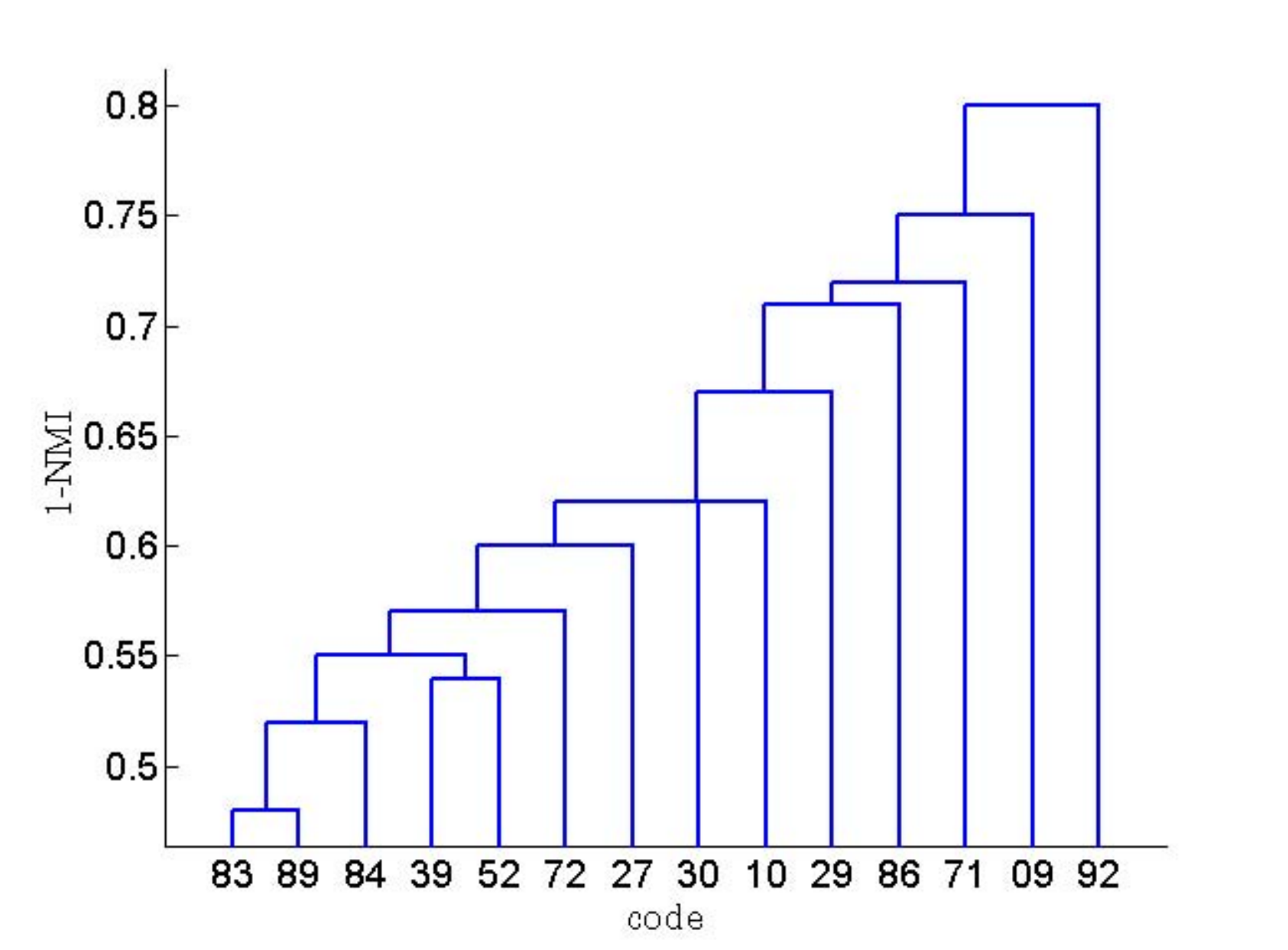}
\caption{{\label{fig:dendro2003}}Minimum spanning tree for 2003. The minimum spanning tree (MST) is computed starting from the similarity between commodity $i$ and commodity $j$ expressed by the index $\mbox{NMI}(i,j)$ and defining a distance between two community structures as $1-\mbox{NMI}(i,j)$. Codes (from left to right): 83 = nuclear reactors; 89 = optical instruments; 84 = electric machinery; 39 = plastics; 52 = cotton; 72 = iron and steel; 27 = mineral fuels; 30 = pharmaceutical products; 10 = cereals; 29 = organic chemicals; 86 = vehicles; 71 = precious stones; 09 = coffee and tea; 92 = arms.}
\end{center}
\end{figure}

The overall similarity pattern between commodity-specific partitions can also be studied using minimum spanning tree (MST) techniques. Indeed, starting from the similarity between commodity $i$ and commodity $j$ expressed by the index $\mbox{NMI}(i,j)$, we can define a distance between two community structures as $1-\mbox{NMI}(i,j)$. Using this metrics, we can build a MST as in Ref. \citep{Mantegna1999} in order to classify the commodities into groups displaying the highest similarity in terms of their community structure. The results for year 2003 is in Figure \ref{fig:dendro2003} (similar results hold for all the remaining years). The figure shows that commodities related to science- or technology-based industries (nuclear reactors, optical instruments, electric machinery) are the most similar in terms of their community structures, whereas arms appears to be the most dissimilar one. More generally, this exercise shows that it is possible to find meaningful classifications of commodities using indicators assessing the similarity between community structures characterizing commodity-specific trade networks.

\subsection{Community Structure, Geography, and Trade Agreements}
We finally turn to study the extent to which community structures identified using trade data correlate with other economics-relevant data. To do that, we employ the geographical closeness matrix $S$, as well the time-dependent RTA networks $M_t$. The entries of the symmetric and time-independent matrix $S$, to repeat, express a measure of geographical closeness between pair of countries, computed as the inverse of geographical distance between their most populated cities. The entries of the symmetric but time-dependent matrices $M_t$ contain, in a give year, the number of trade agreements currently in place between any two countries, irrespective of the type of RTA signed (bilateral, multilateral, commodity- specific, etc.). The underlying assumption is that the higher this number, the closer the two countries are in the RTA space (and thus, according to empirical findings, the larger their expected trade flows). 

\begin{figure}[t]
\begin{center}
\subfigure[RTAs.]{\includegraphics[width=7cm, keepaspectratio=true]{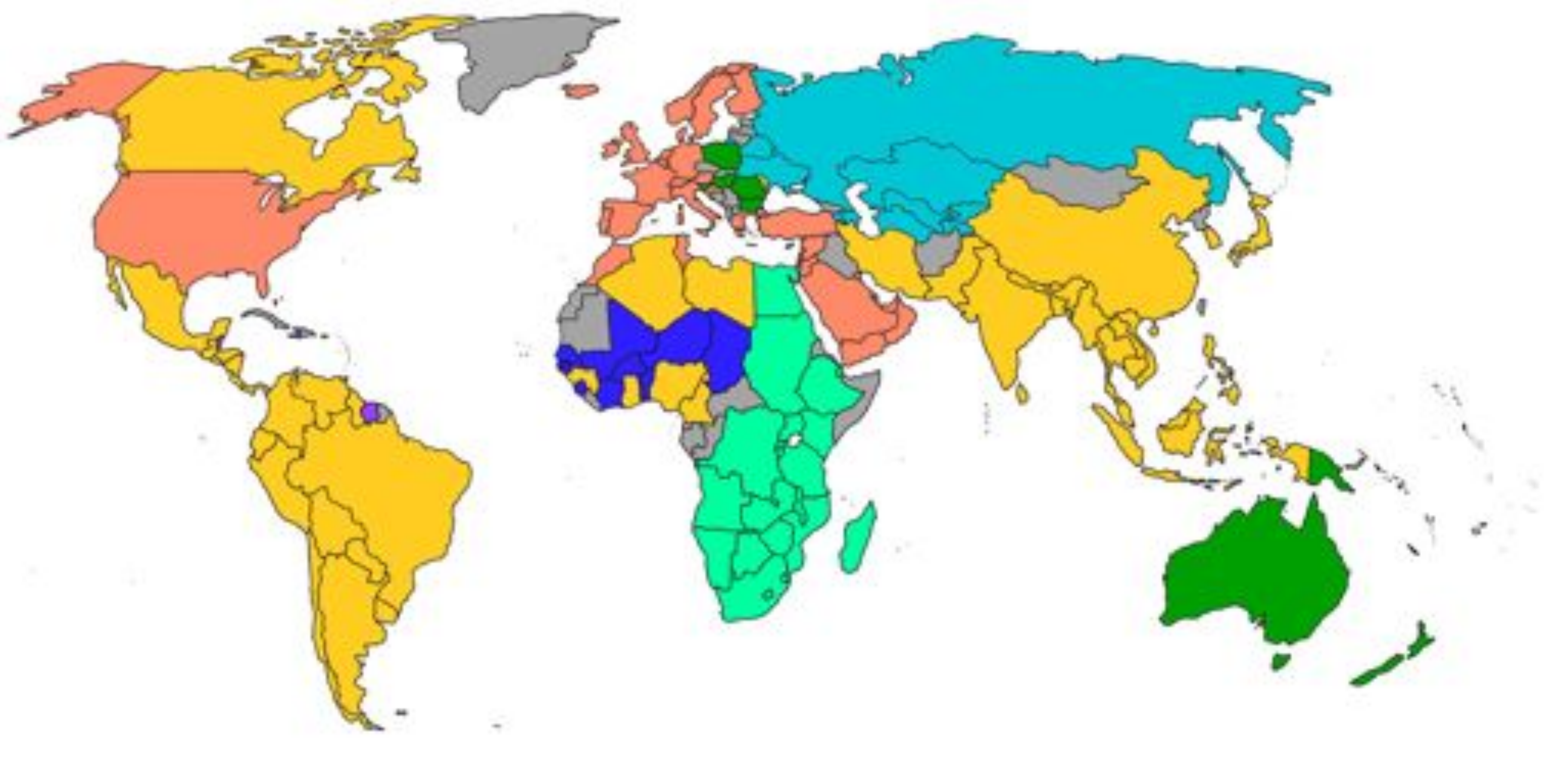}}
\subfigure[Distances.]{\includegraphics[width=7cm, keepaspectratio=true]{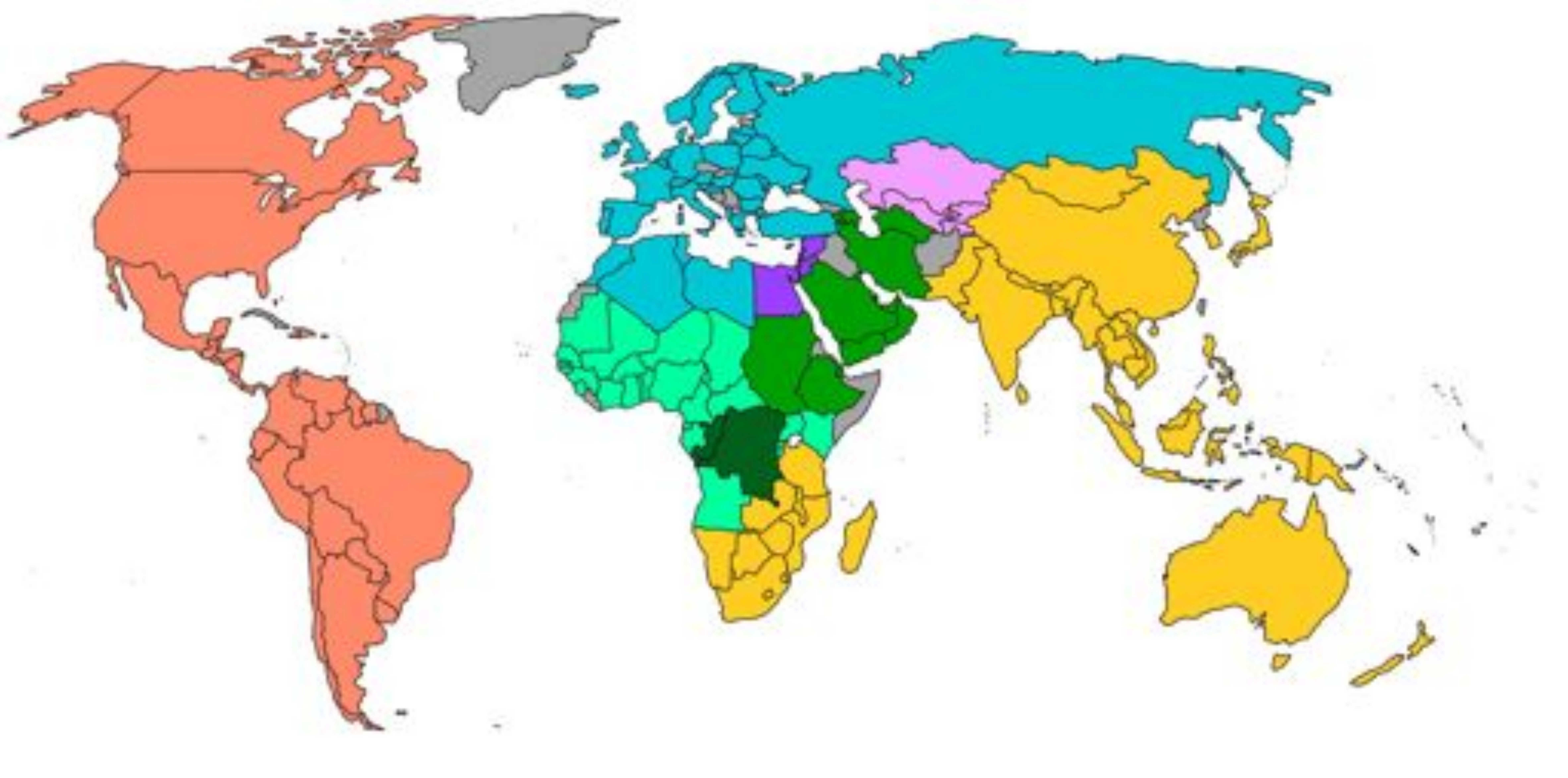}}
\caption{{\label{fig:maps}}World maps showing RTAs in 2003 and geographic communities.}
\end{center}
\end{figure}

We apply to both $S$ and $M_t$ the community-detection algorithms explained in Section \ref{Sec:Comms} and previously applied to trade matrices. Therefore, we end up with a geographically-induced community partition $\mathcal{P}^{GEO}$ and 12 time-dependent RTA-induced partitions $\mathcal{P}_{t}^{RTA}$. The resulting partitions are visualized in the maps of Figure \ref{fig:maps}. Notice that clusters in $\mathcal{P}^{GEO}$ represent groups of countries that are geographically close, without using exogenously-determined partitions of countries (e.g., based on continents or sub-continental breakdowns). The community structures in $\mathcal{P}_{t}^{RTA}$ do instead pick up clusters of countries that not only belong to free-trade or multilateral agreements (e.g. NAFTA, Mercosur, EU, etc.), but also signed additional bilateral agreements.

\begin{figure}[t]
\begin{center}
\includegraphics[width=12cm,keepaspectratio=true]{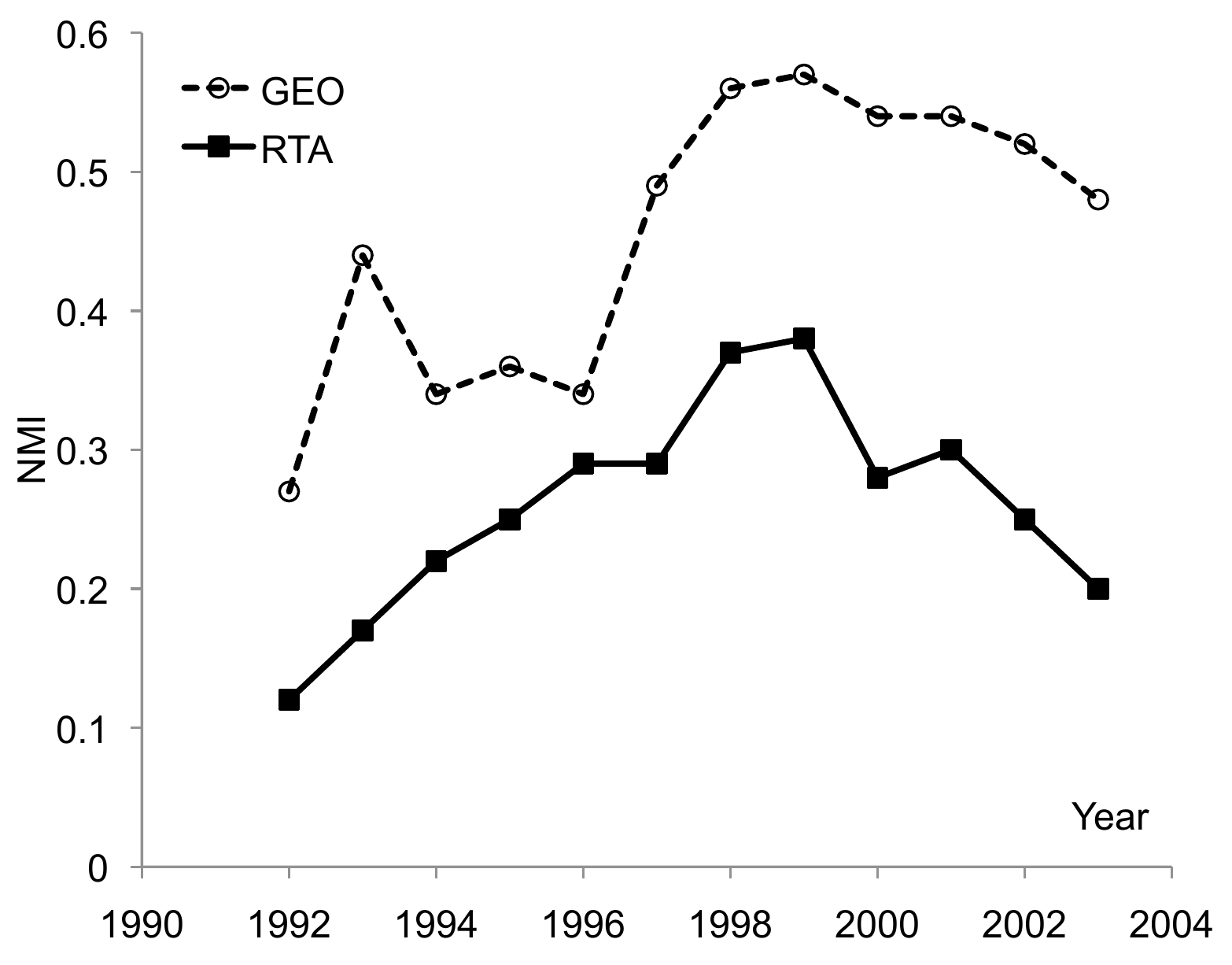}
\caption{NMI when comparing the community structures induced by the exogenous networks build using geographical distances (GEO) or regional trade agreements data (RTA) with the the community structures of aggregate trade.} \label{Fig:NMI_agg}
\end{center}
\end{figure}

We first compare, using the NMI, the aggregate ITN community structure with those detected using geographical distances or RTAs (see Figure \ref{Fig:NMI_agg}). We observe increasing NMIs across time until 2001 and a slight decrease afterwards. We also find more similarity between aggregate trade and geography based communities with respect to communities determined by RTAs. Thus, geographically-related factors seem to explain the pattern of global trade more than political determinants. Also, this result is more evident in the recent years after 2001. A possible explanation might be the global political crisis after 11th September 2001 that implied a slight decrease in global trade as a consequence of the wars in Iraq and Afghanistan.

When comparing community structures of commodity-specific ITNs with the partitions obtained from geographical and RTA data  (see Tables \ref{Tab:NMI_geo_disagg} and \ref{Tab:NMI_fta_disagg}), we find results similar to the aggregate case. In general, it is geography and not trade agreements that seems to correlate more with the observed patterns. Plastics ($c=39$) and mineral fuels ($c=27$) display the highest similarity with RTA communities. The same result holds when confronting trade communities with geographical data, but in addition we notice high NMIs also for iron and steel ($c=72$) and cotton ($c=52$). 

These results reinforce the traditional view put forth by standard gravity-equation trade empirics \cite{Fagiolo2010Jeic}, which stresses the importance of geographical distance (as a proxy for trade resistance factors) in determining bilateral trade flows. Here, we show that geographical distance is important to predict not only the expected flow of a bilateral trade relationship (e.g., exports from country A to country B), but also the formation of trade communities, that is complicated trade structures multilaterally involving groups of countries. On the other hand, our findings contribute to the discussion related to the impact of international agreements on world trade and seem to go in the direction of Ref. \citep{Rose2004}, which shows that there is no evidence that the WTO has increased international trade. 

\section{Concluding Remarks}\label{Sec:Conclusions}
In this paper, we provide a first exploratory study of the community structure of commodity-specific trade networks from 1992-2003. After recovering the optimal partition of countries, we compare commodity-specific communities with the aggregate trade community. 

Our results show that commodity-specific community structures are very heterogeneous and in general their statistical properties are quite different from those of the community structure of the aggregate ITN. For example, whereas the number of communities of the aggregate ITN increases in time, this is not the case for most commodity-specific trade networks. Moreover, the shapes and ranges of cluster-size distributions vary a lot across commodities. As far as community structure evolution is concerned, one observes a decrease in concentration over time of cluster-size distributions (a sign of the globalization process) for the aggregate ITN, a pattern that is not always matched at the commodity-specific level, where trade associated to some products have become less and less centralized and increasingly occurred among smaller and more dispersed groups of countries. Furthermore, the community structure of the aggregate ITN has been changing more slowly over time than their commodity-specific counterparts. 

We have also explored to what extent the community structure of any commodity-specific network may contribute to shape (or is able to predict) that observed at the aggregate-trade level. We have shown that commodity-specific community structures are becoming more and more similar to the aggregate one, i.e. that the role of all commodities in shaping the aggregated community structure has increased in time. However, a major role for the chemical sector appears from these results, in that they are the ones whose country partition better mimics the aggregate one. More generally, by comparing the commodity-specific community structures using the NMI indicator, one can find meaningful classifications of commodities.

Finally, we have explored two possible factors that correlate with community structure, namely geographical distance and the existence of regional trade agreements between countries. Our findings suggest that geographical distance correlates much more with the observed community structure than RTAs. This result confirms previous findings from the empirical literature on trade.

The paper can be extended and refined in at least three directions. First, our findings related to the impact of geography and RTAs are only partial, as they only check for unconditional effects (i.e. they do not address the residual effects of trade agreements once geography is controlled for). In order to make our statements more robust, one may follow Ref. \citep{Reyes_Comm} and compare communities observed in trade data with those detected in the network built with the predictions of a standard gravity model \citep[see for example Ref.][]{Fagiolo2010Jeic} . 

Second, the robustness of our results should be checked against a number of possible problems. For instance, it is well-known that modularity-based community detection suffers from a resolution limit bias \cite{FortunatoBarthelemy07_ModularityResolution}. Therefore, community detection algorithms based on alternative criteria may be employed (e.g., community detection methods based on information theory, see for example Ref. \cite{Rosvall_Bergstrom_2008}). Similarly, one may consider to apply algorithms allowing for overlapping communities \cite{Palla_etal_2005,Nicosia_etal_overlapping}.

Third, another point that deserves further analysis is the detection of community structures across commodity-specific layers. In the paper, we have analyzed independently the most important 14 layers. This allows one to identify groups of countries that trade the same commodity among them. From an economic point of view this signals strong interdependencies but does not convey any insights on the input-output structure of the cluster. For example, there might be groups of countries that are linked in tightly connected chains or cycles, where a country imports from another one a particular type of commodity needed as input for its peculiar industrial structure, and at the same time exports to other countries in the group another commodity that is fed into their production processes (or consumed as final good).  In order to address these issues, one would like to either synthesize into a meaningful statistic all commodity-specific relationships between any two countries or apply new techniques able to detect community structures in multi graphs \citep{multiplex}.



\newpage
\singlespacing

\begin{table}
\centering \scriptsize
\begin{tabular}{p{1cm}p{6.5cm}p{2cm}p{2cm}p{2cm}}
\hline
\hline
 \\
  Code &  Commodity & Value  & Value per Link & $\%$ of Aggregate \\
		&			& (USD)	  &  (USD)		   & Trade\\
\\
\hline
 \\
       83 & Nuclear reactors, boilers, machinery and &   $5.67\times 10^{11}$ &   $6.17\times 10^{7}$ &    11.37\% \\
		   &  mechanical appliances; parts thereof &\\
		   \\
        84 & Electric machinery, equipment and parts;  &   $5.58\times 10^{11}$ &   $6.37\times 10^{7}$ &    11.18\% \\
		   &  sound equipment; television equipment &\\
		   \\
        27 & Mineral fuels, mineral oils \& products of their  &   $4.45\times 10^{11}$ &   $9.91\times 10^{7}$ &     8.92\% \\
			& distillation; bitumin substances; mineral wax  & \\
        \\
		86 & Vehicles (not railway, tramway, rolling stock);$\quad$  &   $3.09\times 10^{11}$ &   $4.76\times 10^{7}$ &     6.19\% \\
		&  parts and accessories&\\
\\
        89 & Optical, photographic, cinematographic, &   $1.78\times 10^{11}$ &   $2.48\times 10^{7}$ &     3.58\% \\
			& measuring, checking, precision, medical or & \\
			& surgical instruments/apparatus;  &\\
			& parts \& accessories &\\
  \\
      39 & Plastics and articles thereof  &   $1.71\times 10^{11}$ &   $2.33\times 10^{7}$ &     3.44\% \\
\\
        29 & Organic chemicals  &   $1.67\times 10^{11}$ &   $3.29\times 10^{7}$ &     3.35\% \\
\\
        30 & Pharmaceutical products  &    $1.4\times 10^{11}$ &   $2.59\times 10^{7}$ &     2.81\% \\
\\
        72 & Iron and steel  &   $1.35\times 10^{11}$ &   $2.77\times 10^{7}$ &     2.70\% \\
\\
        71 & Pearls, precious stones, metals, coins, etc &   $1.01\times 10^{11}$ &   $2.41\times 10^{7}$ &     2.02\% \\
\\
        10 &   Cereals  &   $3.63\times 10^{10}$ &   $1.28\times 10^{7}$ &     0.73\% \\
\\
        52 & Cotton, including yarn and woven fabric thereof  &   $3.29\times 10^{10}$ &   $6.96\times 10^{6}$ &     0.66\% \\
\\
         09 & Coffee, tea, mate \& spices  &   $1.28\times 10^{10}$ &   $2.56\times 10^{6}$ &     0.26\% \\
\\
        92 & Arms and ammunition, parts and  &   $4.31\times 10^{9}$ &   $2.46\times 10^{6}$ &     0.09\% \\
			& accessories thereof &\\
\\
       ALL &  Aggregate &   $4.99\times 10^{12}$ &   $3.54\times 10^{8}$ &   100.00\% \\
\\
\hline
\hline
\end{tabular}
\caption{The 14 most relevant commodity classes in year 2003 in terms of
total-trade value (USD), trade value per link (USD), and share of world
aggregate trade.} \label{Tab:Top14}
\end{table}

\newpage
\begin{sidewaystable}
\center \scriptsize
\begin{tabular}{p{2cm}p{1.5cm}p{1.2cm}p{1.2cm}p{1.2cm}p{1.2cm}p{1.2cm}p{1.2cm}p{1.2cm}p{1.2cm}p{1.2cm}p{1.2cm}p{1.2cm}p{1.2cm}p{1.2cm}p{1.2cm}}
\hline
\hline
\\
Commodity & All &	Coffee	&	Cereals	&	Mineral	&	Organic	&	Pharma. 	&	Plastics	&	Cotton	&	Precious	&	Iron	&	Nuclear	&	Electric	&	Vehicles	&	Optical	&	Arms	\\
 & &	tea	&	&	fuels	&	chem.	&	prod.	&	&	& stones	&steel	&	 reac.	&	 mach.	&		&	inst.	&	\\	
Year&&&&&&&&&&&&&&\\
\\
\hline
\\
1992	&	0.2260	&	38\%	&	17\%	&	30\%	&	36\%	&	33\%	&	44\%	&	36\%	&	34\%	&	33\%	&	58\%	&	55\%	&	42\%	&	45\%	&	14\%	\\
1993	&	0.2832	&	37\%	&	16\%	&	29\%	&	36\%	&	33\%	&	45\%	&	36\%	&	34\%	&	32\%	&	59\%	&	56\%	&	42\%	&	45\%	&	13\%	\\
1994	&	0.3602	&	36\%	&	18\%	&	30\%	&	36\%	&	34\%	&	47\%	&	38\%	&	33\%	&	32\%	&	60\%	&	57\%	&	43\%	&	47\%	&	14\%	\\
1995	&	 0.4199 	&	34\%	&	18\%	&	30\%	&	35\%	&	36\%	&	46\%	&	36\%	&	30\%	&	33\%	&	60\%	&	57\%	&	44\%	&	47\%	&	13\%	\\
1996	&	0.4553	&	35\%	&	19\%	&	30\%	&	35\%	&	35\%	&	47\%	&	35\%	&	29\%	&	33\%	&	61\%	&	58\%	&	45\%	&	46\%	&	13\%	\\
1997	&	0.4925	&	33\%	&	19\%	&	30\%	&	35\%	&	36\%	&	48\%	&	34\%	&	29\%	&	33\%	&	61\%	&	59\%	&	45\%	&	47\%	&	13\%	\\
1998	&	0.5118	&	34\%	&	20\%	&	30\%	&	35\%	&	36\%	&	48\%	&	34\%	&	28\%	&	33\%	&	62\%	&	59\%	&	45\%	&	48\%	&	12\%	\\
1999	&	0.5297	&	33\%	&	20\%	&	30\%	&	35\%	&	37\%	&	49\%	&	34\%	&	28\%	&	33\%	&	62\%	&	59\%	&	45\%	&	47\%	&	12\%	\\
2000	&	0.5406	&	33\%	&	19\%	&	30\%	&	35\%	&	37\%	&	50\%	&	34\%	&	28\%	&	33\%	&	63\%	&	60\%	&	45\%	&	48\%	&	12\%	\\
2001	&	0.5570	&	33\%	&	19\%	&	31\%	&	35\%	&	38\%	&	50\%	&	33\%	&	27\%	&	33\%	&	63\%	&	60\%	&	45\%	&	48\%	&	12\%	\\
2002	&	0.5334	&	33\%	&	20\%	&	32\%	&	35\%	&	38\%	&	51\%	&	33\%	&	28\%	&	33\%	&	64\%	&	61\%	&	46\%	&	49\%	&	12\%	\\
2003	&	0.5400	&	35\%	&	20\%	&	32\%	&	36\%	&	38\%	&	52\%	&	34\%	&	30\%	&	35\%	&	65\%	&	62\%	&	46\%	&	51\%	&	12\%	\\
\\
\hline
\hline
\end{tabular}
\caption{Density of aggregate ITN and relative density of commodity specific ITNs with respect to the aggregate.} \label{Tab:dens}

\center \scriptsize
\begin{tabular}{p{2cm}p{1.2cm}p{1.2cm}p{1.2cm}p{1.2cm}p{1.2cm}p{1.2cm}p{1.2cm}p{1.2cm}p{1.2cm}p{1.2cm}p{1.2cm}p{1.2cm}p{1.2cm}p{1.2cm}p{1.2cm}}
\hline
\hline
\\
Commodity & All &	Coffee	&	Cereals	&	Mineral	&	Organic	&	Pharma. 	&	Plastics	&	Cotton	&	Precious	&	Iron	&	Nuclear	&	Electric	&	Vehicles	&	Optical	&	Arms	\\
 & &	tea	&	&	fuels	&	chem.	&	prod.	&	&	& stones	&steel	&	 reac.	&	 mach.	&		&	inst.	&	\\	
Year&&&&&&&&&&&&&&\\
\\
\hline
\\
1992	&	162	&	145	&	111	&	127	&	138	&	129	&	151	&	147	&	151	&	144	&	161	&	161	&	154	&	157	&	90	\\
1993	&	162	&	150	&	129	&	143	&	143	&	140	&	158	&	153	&	160	&	149	&	162	&	162	&	159	&	162	&	105	\\
1994	&	162	&	155	&	133	&	148	&	152	&	152	&	160	&	156	&	160	&	158	&	162	&	162	&	161	&	160	&	111	\\
1995	&	162	&	158	&	144	&	156	&	157	&	155	&	161	&	156	&	160	&	162	&	162	&	162	&	161	&	161	&	120	\\
1996	&	162	&	158	&	145	&	156	&	153	&	153	&	161	&	157	&	158	&	158	&	162	&	162	&	159	&	162	&	129	\\
1997	&	162	&	162	&	150	&	155	&	151	&	154	&	159	&	156	&	161	&	160	&	162	&	162	&	161	&	162	&	130	\\
1998	&	162	&	161	&	151	&	156	&	157	&	152	&	160	&	157	&	160	&	158	&	162	&	162	&	160	&	162	&	132	\\
1999	&	162	&	160	&	153	&	160	&	156	&	157	&	161	&	159	&	161	&	160	&	162	&	162	&	162	&	162	&	129	\\
2000	&	162	&	160	&	150	&	157	&	154	&	157	&	162	&	160	&	158	&	161	&	162	&	162	&	161	&	162	&	137	\\
2001	&	162	&	161	&	152	&	160	&	160	&	160	&	161	&	157	&	159	&	160	&	162	&	162	&	162	&	162	&	139	\\
2002	&	162	&	160	&	150	&	160	&	158	&	158	&	161	&	158	&	158	&	157	&	162	&	162	&	162	&	162	&	139	\\
2003	&	162	&	161	&	150	&	158	&	157	&	158	&	162	&	161	&	161	&	159	&	162	&	162	&	162	&	162	&	134	\\
\\
\hline
\hline
\end{tabular}
\caption{Size of the largest connected components (LCC).} \label{Tab:greatconn}
\end{sidewaystable}

\begin{sidewaystable}
\center \scriptsize
\begin{tabular}{p{2cm}p{1.2cm}p{1.2cm}p{1.2cm}p{1.2cm}p{1.2cm}p{1.2cm}p{1.2cm}p{1.2cm}p{1.2cm}p{1.2cm}p{1.2cm}p{1.2cm}p{1.2cm}p{1.2cm}p{1.2cm}}
\hline
\hline
\\
Commodity & All &	Coffee	&	Cereals	&	Mineral	&	Organic	&	Pharma. 	&	Plastics	&	Cotton	&	Precious	&	Iron	&	Nuclear	&	Electric	&	Vehicles	&	Optical	&	Arms	\\
 & &	tea	&	&	fuels	&	chem.	&	prod.	&	&	& stones	&steel	&	 reac.	&	 mach.	&		&	inst.	&	\\	
   Year&&&&&&&&&&&&&&\\
\\
\hline
\\
1992	&	2	&	6	&	5	&	4	&	3	&	6	&	3	&	5	&	5	&	3	&	3	&	4	&	3	&	2	&	7	\\
1993	&	3	&	10	&	5	&	4	&	3	&	8	&	3	&	6	&	5	&	6	&	4	&	3	&	4	&	2	&	5	\\
1994	&	3	&	8	&	8	&	5	&	3	&	6	&	4	&	4	&	5	&	5	&	3	&	2	&	4	&	2	&	6	\\
1995	&	3	&	7	&	7	&	8	&	3	&	8	&	7	&	5	&	5	&	5	&	5	&	4	&	5	&	4	&	8	\\
1996	&	3	&	5	&	7	&	6	&	4	&	6	&	5	&	6	&	5	&	6	&	4	&	3	&	5	&	3	&	9	\\
1997	&	4	&	6	&	6	&	6	&	6	&	6	&	4	&	5	&	5	&	4	&	4	&	2	&	6	&	2	&	6	\\
1998	&	4	&	6	&	6	&	6	&	4	&	6	&	6	&	5	&	4	&	4	&	5	&	5	&	4	&	2	&	7	\\
1999	&	4	&	6	&	8	&	8	&	3	&	5	&	5	&	5	&	4	&	4	&	5	&	4	&	5	&	3	&	8	\\
2000	&	4	&	7	&	6	&	5	&	4	&	6	&	6	&	6	&	4	&	4	&	5	&	4	&	6	&	3	&	6	\\
2001	&	4	&	6	&	5	&	5	&	3	&	7	&	5	&	6	&	4	&	5	&	5	&	3	&	4	&	3	&	6	\\
2002	&	4	&	6	&	6	&	4	&	4	&	6	&	7	&	5	&	4	&	5	&	2	&	3	&	5	&	3	&	8	\\
2003	&	4	&	7	&	6	&	6	&	4	&	6	&	6	&	6	&	5	&	4	&	5	&	2	&	5	&	3	&	9	\\
\\
\hline
\hline
\end{tabular}
\caption{Number of communities.} \label{Tab:numcomp}

\center \scriptsize
\begin{tabular}{p{2cm}p{1.2cm}p{1.2cm}p{1.2cm}p{1.2cm}p{1.2cm}p{1.2cm}p{1.2cm}p{1.2cm}p{1.2cm}p{1.2cm}p{1.2cm}p{1.2cm}p{1.2cm}p{1.2cm}p{1.2cm}}
\hline
\hline
\\
Commodity & All &	Coffee	&	Cereals	&	Mineral	&	Organic	&	Pharma. 	&	Plastics	&	Cotton	&	Precious	&	Iron	&	Nuclear	&	Electric	&	Vehicles	&	Optical	&	Arms	\\
 & &	tea	&	&	fuels	&	chem.	&	prod.	&	&	& stones	&steel	&	 reac.	&	 mach.	&		&	inst.	&	\\	
   Year&&&&&&&&&&&&&&\\
\\
\hline
\\
1992	&	0.50	&	0.16	&	0.12	&	0.16	&	0.25	&	0.14	&	0.30	&	0.30	&	0.30	&	0.27	&	0.38	&	0.41	&	0.49	&	0.48	&	0.08	\\
1993	&	0.34	&	0.11	&	0.16	&	0.21	&	0.26	&	0.13	&	0.34	&	0.34	&	0.33	&	0.24	&	0.40	&	0.50	&	0.35	&	0.55	&	0.12	\\
1994	&	0.46	&	0.15	&	0.14	&	0.26	&	0.36	&	0.18	&	0.33	&	0.33	&	0.26	&	0.30	&	0.44	&	0.51	&	0.35	&	0.52	&	0.16	\\
1995	&	0.44	&	0.17	&	0.20	&	0.22	&	0.46	&	0.15	&	0.24	&	0.24	&	0.28	&	0.25	&	0.34	&	0.45	&	0.41	&	0.47	&	0.11	\\
1996	&	0.41	&	0.24	&	0.16	&	0.20	&	0.38	&	0.18	&	0.37	&	0.37	&	0.27	&	0.28	&	0.35	&	0.40	&	0.44	&	0.45	&	0.14	\\
1997	&	0.29	&	0.20	&	0.17	&	0.24	&	0.31	&	0.21	&	0.29	&	0.29	&	0.35	&	0.25	&	0.36	&	0.53	&	0.43	&	0.56	&	0.20	\\
1998	&	0.29	&	0.19	&	0.19	&	0.21	&	0.30	&	0.18	&	0.25	&	0.25	&	0.37	&	0.25	&	0.24	&	0.48	&	0.45	&	0.54	&	0.16	\\
1999	&	0.30	&	0.20	&	0.18	&	0.23	&	0.32	&	0.25	&	0.27	&	0.27	&	0.30	&	0.25	&	0.37	&	0.40	&	0.39	&	0.44	&	0.13	\\
2000	&	0.28	&	0.22	&	0.20	&	0.22	&	0.25	&	0.24	&	0.27	&	0.27	&	0.28	&	0.26	&	0.34	&	0.44	&	0.36	&	0.39	&	0.20	\\
2001	&	0.28	&	0.23	&	0.22	&	0.27	&	0.34	&	0.18	&	0.26	&	0.26	&	0.31	&	0.22	&	0.35	&	0.45	&	0.36	&	0.46	&	0.23	\\
2002	&	0.29	&	0.24	&	0.21	&	0.27	&	0.30	&	0.25	&	0.24	&	0.24	&	0.28	&	0.21	&	0.53	&	0.41	&	0.46	&	0.45	&	0.15	\\
2003	&	0.31	&	0.23	&	0.16	&	0.24	&	0.26	&	0.19	&	0.20	&	0.20	&	0.27	&	0.26	&	0.37	&	0.53	&	0.41	&	0.45	&	0.13	\\
\\
\hline
\hline
\end{tabular}
\caption{Normalized Herfindal Index $\mbox{H}_t^{c}$ as a measure of concentration of communities.} \label{Tab:HI}
\end{sidewaystable}

\begin{sidewaystable}
\center \scriptsize
\begin{tabular}{p{2cm}p{1.2cm}p{1.2cm}p{1.2cm}p{1.2cm}p{1.2cm}p{1.2cm}p{1.2cm}p{1.2cm}p{1.2cm}p{1.2cm}p{1.2cm}p{1.2cm}p{1.2cm}p{1.2cm}p{1.2cm}}
\hline
\hline
\\
Commodity & All &	Coffee	&	Cereals	&	Mineral	&	Organic	&	Pharma. 	&	Plastics	&	Cotton	&	Precious	&	Iron	&	Nuclear	&	Electric	&	Vehicles	&	Optical	&	Arms	\\
 & &	tea	&	&	fuels	&	chem.	&	prod.	&	&	& stones	&steel	&	 reac.	&	 mach.	&		&	inst.	&	\\	
   Year&&&&&&&&&&&&&&\\
\\
\hline
\\
1992	-	2003	&	0.27	&	0.17	&	0.22	&	0.29	&	0.24	&	0.17	&	0.35	&	0.28	&	0.20	&	0.29	&	0.23	&	0.22	&	0.20	&	0.30	&	0.16	\\
\\
1992	-	1993	&	0.54	&	0.38	&	0.25	&	0.35	&	0.27	&	0.28	&	0.42	&	0.29	&	0.35	&	0.39	&	0.38	&	0.20	&	0.25	&	0.25	&	0.33	\\
1993	-	1994	&	0.41	&	0.46	&	0.32	&	0.48	&	0.24	&	0.35	&	0.48	&	0.34	&	0.32	&	0.41	&	0.38	&	0.48	&	0.42	&	0.37	&	0.35	\\
1994	-	1995	&	0.55	&	0.42	&	0.43	&	0.50	&	0.28	&	0.45	&	0.45	&	0.45	&	0.17	&	0.45	&	0.37	&	0.51	&	0.29	&	0.42	&	0.29	\\
1995	-	1996	&	0.51	&	0.33	&	0.47	&	0.52	&	0.37	&	0.44	&	0.52	&	0.34	&	0.27	&	0.46	&	0.40	&	0.56	&	0.51	&	0.56	&	0.33	\\
1996	-	1997	&	0.58	&	0.27	&	0.34	&	0.50	&	0.40	&	0.56	&	0.39	&	0.35	&	0.43	&	0.46	&	0.40	&	0.55	&	0.52	&	0.50	&	0.32	\\
1997	-	1998	&	0.75	&	0.51	&	0.43	&	0.61	&	0.32	&	0.51	&	0.56	&	0.44	&	0.48	&	0.56	&	0.54	&	0.59	&	0.58	&	0.71	&	0.33	\\
1998	-	1999	&	0.77	&	0.56	&	0.52	&	0.71	&	0.56	&	0.48	&	0.57	&	0.35	&	0.42	&	0.57	&	0.39	&	0.47	&	0.51	&	0.39	&	0.34	\\
1999	-	2000	&	0.73	&	0.49	&	0.49	&	0.55	&	0.40	&	0.44	&	0.59	&	0.42	&	0.49	&	0.44	&	0.50	&	0.50	&	0.55	&	0.42	&	0.33	\\
2000	-	2001	&	0.79	&	0.63	&	0.47	&	0.60	&	0.38	&	0.54	&	0.46	&	0.55	&	0.47	&	0.52	&	0.56	&	0.50	&	0.42	&	0.40	&	0.29	\\
2001	-	2002	&	0.68	&	0.55	&	0.55	&	0.59	&	0.34	&	0.64	&	0.45	&	0.41	&	0.34	&	0.57	&	0.43	&	0.44	&	0.41	&	0.63	&	0.30	\\
2002	-	2003	&	0.65	&	0.49	&	0.53	&	0.55	&	0.42	&	0.46	&	0.57	&	0.52	&	0.32	&	0.52	&	0.37	&	0.58	&	0.47	&	0.58	&	0.28	\\
\\
\hline
\hline
\end{tabular}
\caption{NMI when comparing the community structures along the time dimension.} \label{Tab:NMI_time}

\center \scriptsize
\begin{tabular}{p{2cm}p{1.2cm}p{1.2cm}p{1.2cm}p{1.2cm}p{1.2cm}p{1.2cm}p{1.2cm}p{1.2cm}p{1.2cm}p{1.2cm}p{1.2cm}p{1.2cm}p{1.2cm}p{1.2cm}}
\hline
\hline
\\
Commodity & 	Coffee	&	Cereals	&	Mineral	&	Organic	&	Pharma. 	&	Plastics	&	Cotton	&	Precious	&	Iron	&	Nuclear	&	Electric	&	Vehicles	&	Optical	&	Arms	\\
 & 	tea	&	&	fuels	&	chem.	&	prod.	&	&	& stones	&steel	&	 reac.	&	 mach.	&		&	inst.	&	\\	
   Year&&&&&&&&&&&&&\\
\\
\hline
\\
1992	&	0.06	&	0.11	&	0.18	&	0.12	&	0.09	&	0.16	&	0.21	&	0.10	&	0.14	&	0.23	&	0.21	&	0.25	&	0.21	&	0.05	\\
1993	&	0.17	&	0.14	&	0.37	&	0.29	&	0.23	&	0.44	&	0.24	&	0.18	&	0.28	&	0.25	&	0.19	&	0.16	&	0.21	&	0.11	\\
1994	&	0.11	&	0.15	&	0.23	&	0.31	&	0.18	&	0.32	&	0.28	&	0.13	&	0.18	&	0.35	&	0.35	&	0.22	&	0.40	&	0.12	\\
1995	&	0.14	&	0.20	&	0.25	&	0.30	&	0.22	&	0.40	&	0.28	&	0.21	&	0.24	&	0.36	&	0.40	&	0.19	&	0.26	&	0.15	\\
1996	&	0.05	&	0.17	&	0.39	&	0.32	&	0.22	&	0.46	&	0.24	&	0.09	&	0.25	&	0.39	&	0.41	&	0.40	&	0.28	&	0.21	\\
1997	&	0.12	&	0.32	&	0.46	&	0.33	&	0.28	&	0.30	&	0.29	&	0.20	&	0.37	&	0.46	&	0.25	&	0.30	&	0.24	&	0.18	\\
1998	&	0.18	&	0.25	&	0.54	&	0.33	&	0.34	&	0.39	&	0.27	&	0.18	&	0.42	&	0.42	&	0.29	&	0.37	&	0.30	&	0.16	\\
1999	&	0.15	&	0.33	&	0.56	&	0.31	&	0.29	&	0.51	&	0.38	&	0.26	&	0.37	&	0.39	&	0.38	&	0.31	&	0.31	&	0.12	\\
2000	&	0.23	&	0.25	&	0.43	&	0.34	&	0.32	&	0.34	&	0.41	&	0.25	&	0.29	&	0.46	&	0.41	&	0.35	&	0.28	&	0.17	\\
2001	&	0.20	&	0.26	&	0.54	&	0.38	&	0.31	&	0.50	&	0.33	&	0.28	&	0.43	&	0.42	&	0.40	&	0.47	&	0.39	&	0.15	\\
2002	&	0.24	&	0.33	&	0.54	&	0.29	&	0.28	&	0.35	&	0.39	&	0.21	&	0.35	&	0.28	&	0.33	&	0.28	&	0.28	&	0.16	\\
2003	&	0.14	&	0.27	&	0.49	&	0.32	&	0.29	&	0.45	&	0.38	&	0.28	&	0.38	&	0.37	&	0.19	&	0.25	&	0.26	&	0.12	\\
\\
\hline
\hline
\end{tabular}
\caption{NMI when comparing the community structures induced by aggregate ITN with commodity-specific ITNs.} \label{Tab:NMI_agg_disagg}
\end{sidewaystable}

\begin{sidewaystable}
\center \scriptsize
\begin{tabular}{p{2cm}p{1.2cm}p{1.2cm}p{1.2cm}p{1.2cm}p{1.2cm}p{1.2cm}p{1.2cm}p{1.2cm}p{1.2cm}p{1.2cm}p{1.2cm}p{1.2cm}p{1.2cm}p{1.2cm}}
\hline
\hline
\\
Commodity  &	Coffee	&	Cereals	&	Mineral	&	Organic	&	Pharma. 	&	Plastics	&	Cotton	&	Precious	&	Iron	&	Nuclear	&	Electric	&	Vehicles	&	Optical	&	Arms	\\
 & 	tea	&	&	fuels	&	chem.	&	prod.	&	&	& stones	&steel	&	 reac.	&	 mach.	&		&	inst.	&	\\	
   Year&&&&&&&&&&&&&\\
\\
\hline
\\
1992	&	0.24	&	0.26	&	0.41	&	0.31	&	0.29	&	0.37	&	0.30	&	0.21	&	0.30	&	0.25	&	0.24	&	0.29	&	0.26	&	0.21	\\
1993	&	0.24	&	0.28	&	0.38	&	0.37	&	0.38	&	0.43	&	0.32	&	0.25	&	0.42	&	0.31	&	0.24	&	0.28	&	0.21	&	0.22	\\
1994	&	0.23	&	0.38	&	0.47	&	0.27	&	0.34	&	0.43	&	0.30	&	0.21	&	0.36	&	0.29	&	0.28	&	0.38	&	0.27	&	0.23	\\
1995	&	0.26	&	0.41	&	0.52	&	0.30	&	0.33	&	0.54	&	0.27	&	0.31	&	0.43	&	0.34	&	0.31	&	0.31	&	0.31	&	0.30	\\
1996	&	0.18	&	0.32	&	0.43	&	0.26	&	0.31	&	0.38	&	0.32	&	0.22	&	0.43	&	0.30	&	0.34	&	0.32	&	0.33	&	0.28	\\
1997	&	0.23	&	0.42	&	0.48	&	0.34	&	0.31	&	0.42	&	0.40	&	0.27	&	0.52	&	0.43	&	0.31	&	0.31	&	0.27	&	0.26	\\
1998	&	0.21	&	0.37	&	0.52	&	0.39	&	0.39	&	0.43	&	0.35	&	0.22	&	0.41	&	0.41	&	0.31	&	0.35	&	0.31	&	0.23	\\
1999	&	0.20	&	0.43	&	0.56	&	0.30	&	0.29	&	0.47	&	0.48	&	0.32	&	0.44	&	0.42	&	0.44	&	0.26	&	0.36	&	0.21	\\
2000	&	0.30	&	0.36	&	0.45	&	0.40	&	0.40	&	0.44	&	0.48	&	0.33	&	0.37	&	0.43	&	0.43	&	0.33	&	0.30	&	0.21	\\
2001	&	0.26	&	0.39	&	0.52	&	0.33	&	0.36	&	0.53	&	0.44	&	0.35	&	0.46	&	0.38	&	0.45	&	0.38	&	0.41	&	0.17	\\
2002	&	0.27	&	0.41	&	0.53	&	0.31	&	0.32	&	0.47	&	0.44	&	0.25	&	0.47	&	0.30	&	0.35	&	0.33	&	0.32	&	0.21	\\
2003	&	0.20	&	0.39	&	0.50	&	0.33	&	0.37	&	0.57	&	0.44	&	0.29	&	0.52	&	0.36	&	0.27	&	0.33	&	0.32	&	0.19	\\
\\
\hline
\hline
\end{tabular}
\caption{NMI when comparing the community structures induced by geographical distances with commodity-specific ITNs.} \label{Tab:NMI_geo_disagg}

\center \scriptsize
\begin{tabular}{p{2cm}p{1.2cm}p{1.2cm}p{1.2cm}p{1.2cm}p{1.2cm}p{1.2cm}p{1.2cm}p{1.2cm}p{1.2cm}p{1.2cm}p{1.2cm}p{1.2cm}p{1.2cm}p{1.2cm}}
\hline
\hline
\\
Commodity  &	Coffee	&	Cereals	&	Mineral	&	Organic	&	Pharma. 	&	Plastics	&	Cotton	&	Precious	&	Iron	&	Nuclear	&	Electric	&	Vehicles	&	Optical	&	Arms	\\
 & 	tea	&	&	fuels	&	chem.	&	prod.	&	&	& stones	&steel	&	 reac.	&	 mach.	&		&	inst.	&	\\	
   Year&&&&&&&&&&&&&\\
\\
\hline
\\
1992	&	0.15	&	0.17	&	0.20	&	0.13	&	0.20	&	0.17	&	0.22	&	0.12	&	0.19	&	0.16	&	0.15	&	0.14	&	0.13	&	0.15	\\
1993	&	0.19	&	0.19	&	0.21	&	0.13	&	0.22	&	0.15	&	0.24	&	0.13	&	0.23	&	0.19	&	0.14	&	0.15	&	0.12	&	0.15	\\
1994	&	0.16	&	0.27	&	0.24	&	0.21	&	0.26	&	0.21	&	0.19	&	0.14	&	0.26	&	0.16	&	0.20	&	0.19	&	0.16	&	0.17	\\
1995	&	0.18	&	0.29	&	0.37	&	0.18	&	0.32	&	0.38	&	0.23	&	0.18	&	0.31	&	0.30	&	0.27	&	0.25	&	0.24	&	0.20	\\
1996	&	0.11	&	0.28	&	0.33	&	0.24	&	0.27	&	0.31	&	0.26	&	0.16	&	0.26	&	0.25	&	0.30	&	0.29	&	0.21	&	0.23	\\
1997	&	0.13	&	0.33	&	0.30	&	0.27	&	0.24	&	0.26	&	0.24	&	0.15	&	0.22	&	0.28	&	0.17	&	0.23	&	0.17	&	0.14	\\
1998	&	0.27	&	0.31	&	0.36	&	0.22	&	0.36	&	0.42	&	0.28	&	0.14	&	0.27	&	0.32	&	0.25	&	0.30	&	0.20	&	0.21	\\
1999	&	0.19	&	0.33	&	0.37	&	0.20	&	0.32	&	0.41	&	0.30	&	0.23	&	0.31	&	0.24	&	0.24	&	0.25	&	0.22	&	0.21	\\
2000	&	0.25	&	0.29	&	0.35	&	0.22	&	0.25	&	0.33	&	0.33	&	0.22	&	0.26	&	0.26	&	0.30	&	0.24	&	0.15	&	0.17	\\
2001	&	0.20	&	0.25	&	0.31	&	0.18	&	0.27	&	0.34	&	0.27	&	0.21	&	0.28	&	0.29	&	0.26	&	0.25	&	0.22	&	0.17	\\
2002	&	0.20	&	0.30	&	0.30	&	0.19	&	0.25	&	0.34	&	0.31	&	0.20	&	0.29	&	0.17	&	0.23	&	0.25	&	0.22	&	0.19	\\
2003	&	0.16	&	0.24	&	0.32	&	0.20	&	0.29	&	0.33	&	0.24	&	0.16	&	0.24	&	0.25	&	0.21	&	0.26	&	0.19	&	0.18	\\
\\
\hline
\hline
\end{tabular}
\caption{NMI when comparing the community structures induced by regional trade agreements with commodity-specific ITNs.} \label{Tab:NMI_fta_disagg}
\end{sidewaystable}


\clearpage

\end{document}